%% file: gplus-triangles.tex
\documentclass[10pt,conference,compsocconf,letterpaper]{IEEEtran}

\usepackage{cite}

\usepackage[cmex10]{amsmath}
\usepackage{url}

\usepackage{times}
\usepackage{amssymb}
\usepackage{wrapfig}
\usepackage{color, array, colortbl}
\definecolor{shadecolor}{rgb}{0.9,0.9,0.9}
\usepackage{comment}

\usepackage{graphicx}
\usepackage{placeins}
\usepackage{framed}
\usepackage{booktabs}
\usepackage{multirow}
\usepackage{xspace}

\input{utils}

\begin{document}
%

\title{Evolution of Directed Triangle Motifs\\in the Google$+$
    OSN\vspace{-0.5\baselineskip}}



%
\author{
\IEEEauthorblockN{%
    Doris Schi\"oberg\IEEEauthorrefmark{1},
    Fabian Schneider\IEEEauthorrefmark{2},
    Stefan Schmid\IEEEauthorrefmark{1},
    Steve Uhlig\IEEEauthorrefmark{3} and
    Anja Feldmann\IEEEauthorrefmark{1}}
\IEEEauthorblockA{\IEEEauthorrefmark{1}
    TU Berlin, Germany ---
    \texttt{\{doris,stefan,anja\}@net.t-labs.tu-berlin.de}}
\IEEEauthorblockA{\IEEEauthorrefmark{2}
    NEC Laboratories Europe, Heidelberg, Germany ---
    \texttt{fabian@ieee.org}}
\IEEEauthorblockA{\IEEEauthorrefmark{3}
    Queen Mary University, London, United Kingdom ---
    \texttt{steve@eecs.qmul.ac.uk}}
\vspace{0.5\baselineskip}
}


\maketitle

\begin{abstract}
Motifs are a fundamental building block and distinguishing feature of networks.
While characteristic motif distributions have been found in many networks,
very little is known today about the \emph{evolution} of network motifs.

This paper studies the most important motifs in social networks, triangles, and
how directed triangle motifs change over time. Our chosen subject is one of the
largest Online Social Networks, Google$+$. Google$+$ has two distinguishing features
that make it particularly interesting: (1) it is a \emph{directed} network, which
yields a rich set of triangle motifs, and (2) it is a young and fast evolving network,
whose role in the OSN space is still not fully understood.

For the purpose of this study, we crawled the network over a time period of six weeks,
collecting several snapshots. We find that some triangle types display significant dynamics,
e.g., for some specific initial types, up to 20$\%$ of the instances evolve to other types.
Due to the fast growth of the OSN in the observed time period, many new triangles emerge.
We also observe that many triangles evolve into \emph{less-connected} motifs (with less
edges), suggesting that growth also comes with pruning.

We complement the topological study by also considering publicly available user profile
data (mostly geographic locations). The corresponding results shed some light on
the semantics of the triangle motifs. Indeed, we find that users in more symmetric triangle
motifs live closer together, indicating more personal relationships. In contrast,
asymmetric links in motifs often point to faraway users with a high in-degree (``celebrities'').
\end{abstract}


%
\IEEEpeerreviewmaketitle

%

\section{Introduction}

Network \emph{motifs}~\cite{motifs}, also known as ~\emph{graphlets} or
\emph{structural signatures}, can give insights into the relationships and
interaction patterns in a network. The existence and frequency distribution
of network motifs has been analyzed in multiple contexts, including
biological~\cite{freq-bio} (e.g., protein-protein interaction networks),
economical~\cite{motif-economics} (e.g., connectivity during mergers) and
social networks.

Apart from links--that is 2-node motifs--the most simple and important
network motif in social networks is the \emph{triangle}: it describes the
relationship between three nodes.  Triangles give insights into the
inter-connectivity of nodes in graphs~\cite{cite13,watts1998cds} and are an
indication of community behavior~\cite{cikm-tri-deg}. Triangles also form
the basis of the widely-studied clustering
coefficient~\cite{cite11,cite27}.

However, while researchers have made much progress on the characterization
of a \emph{given} network, only very little is known today about the
evolution of motifs \emph{over time}. Moreover, many studies today focus on
undirected motifs, especially in the context of clustering (see also the
notion of clustering coefficient itself); in reality however, networks
often feature rich directed relationships.

\textbf{Our Contribution.}
This paper argues that the motifs of any active network are inherently
dynamic, and the study of the motif changes over time can give interesting
new insights into the nature of a network.

\begin{figure}[t]
\centering
\includegraphics[width=0.17\columnwidth]{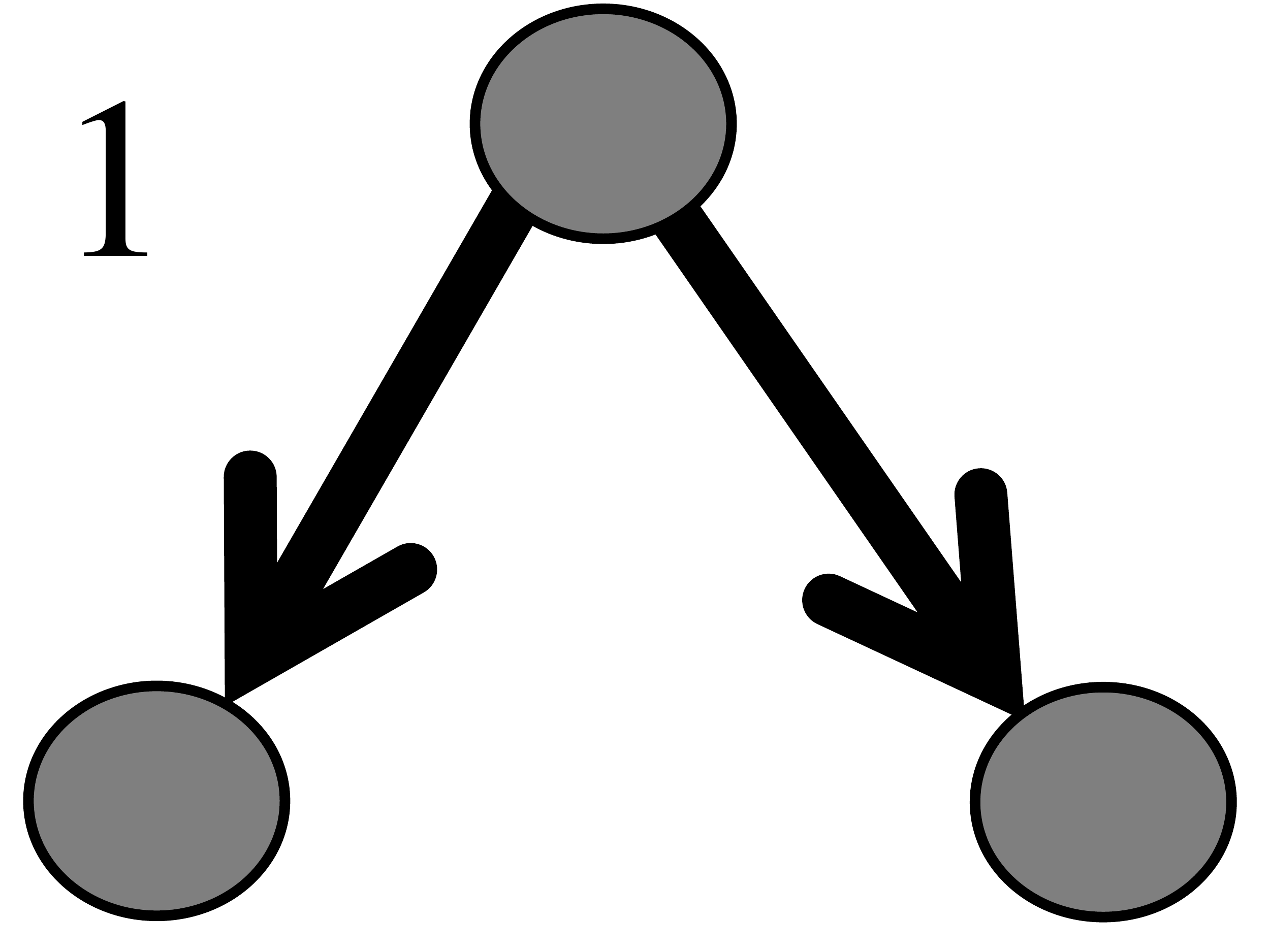}
\includegraphics[width=0.17\columnwidth]{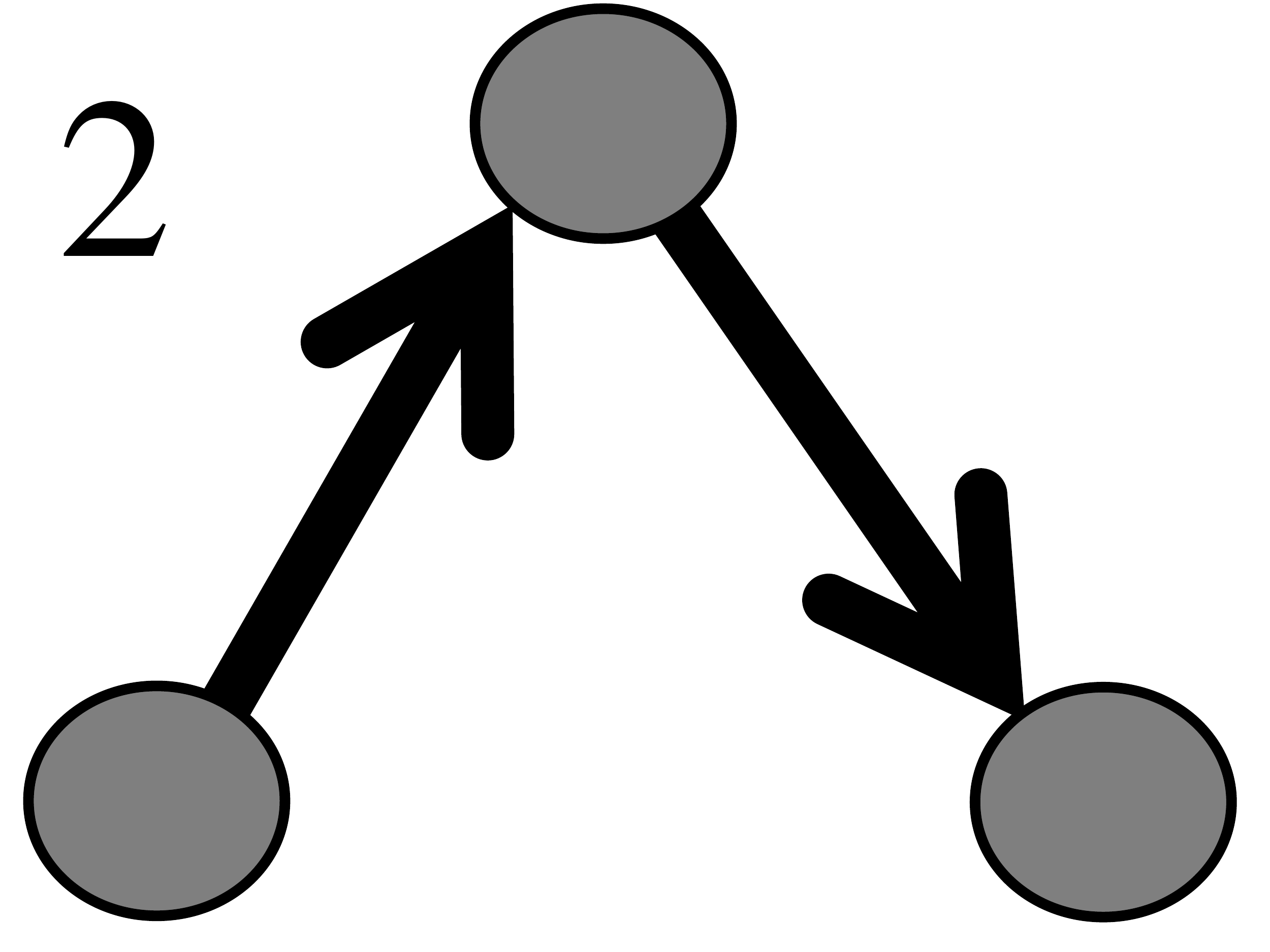}
\includegraphics[width=0.17\columnwidth]{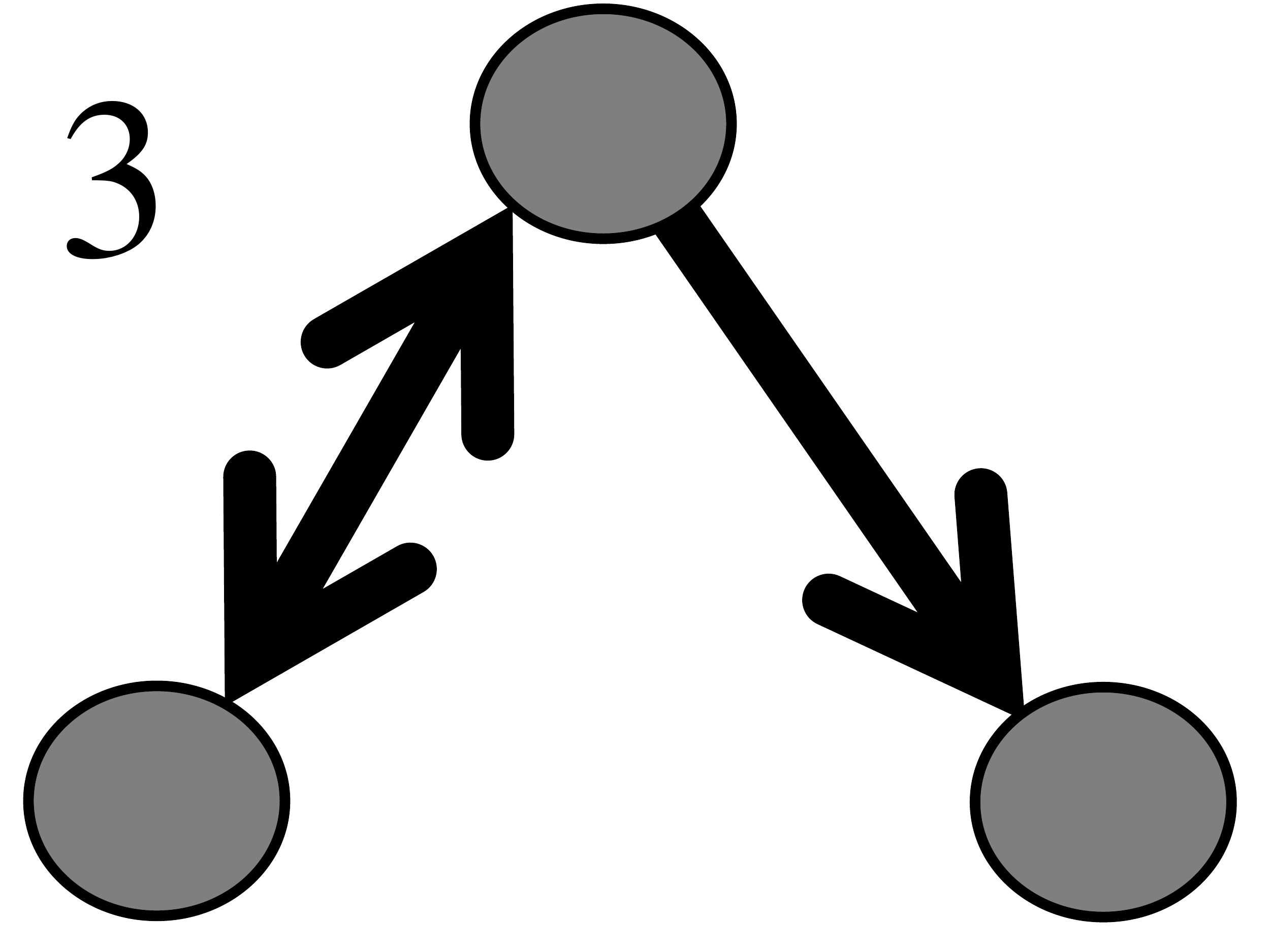}
\includegraphics[width=0.17\columnwidth]{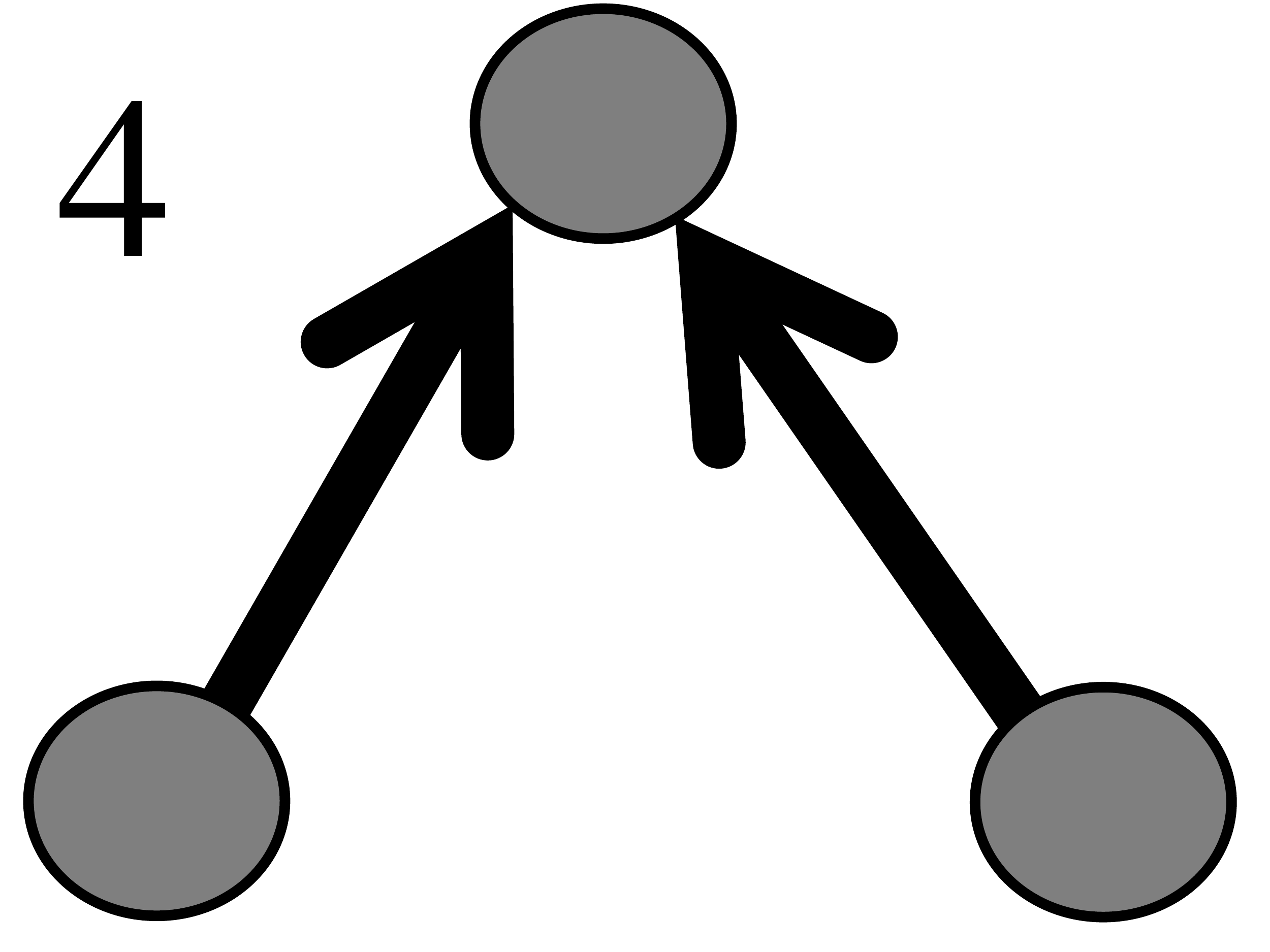}\\ \vspace{2mm}
\includegraphics[width=0.17\columnwidth]{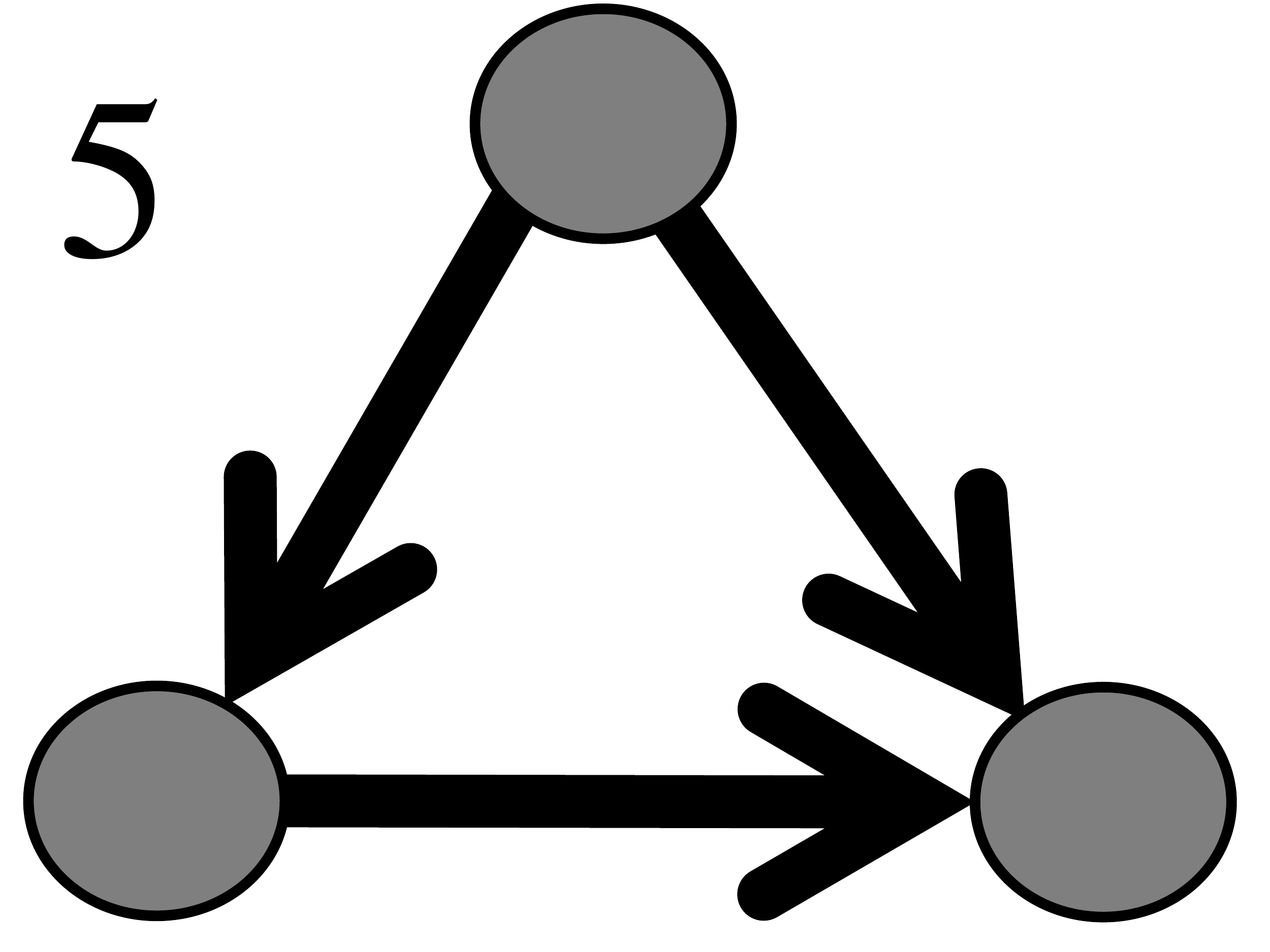}
\includegraphics[width=0.17\columnwidth]{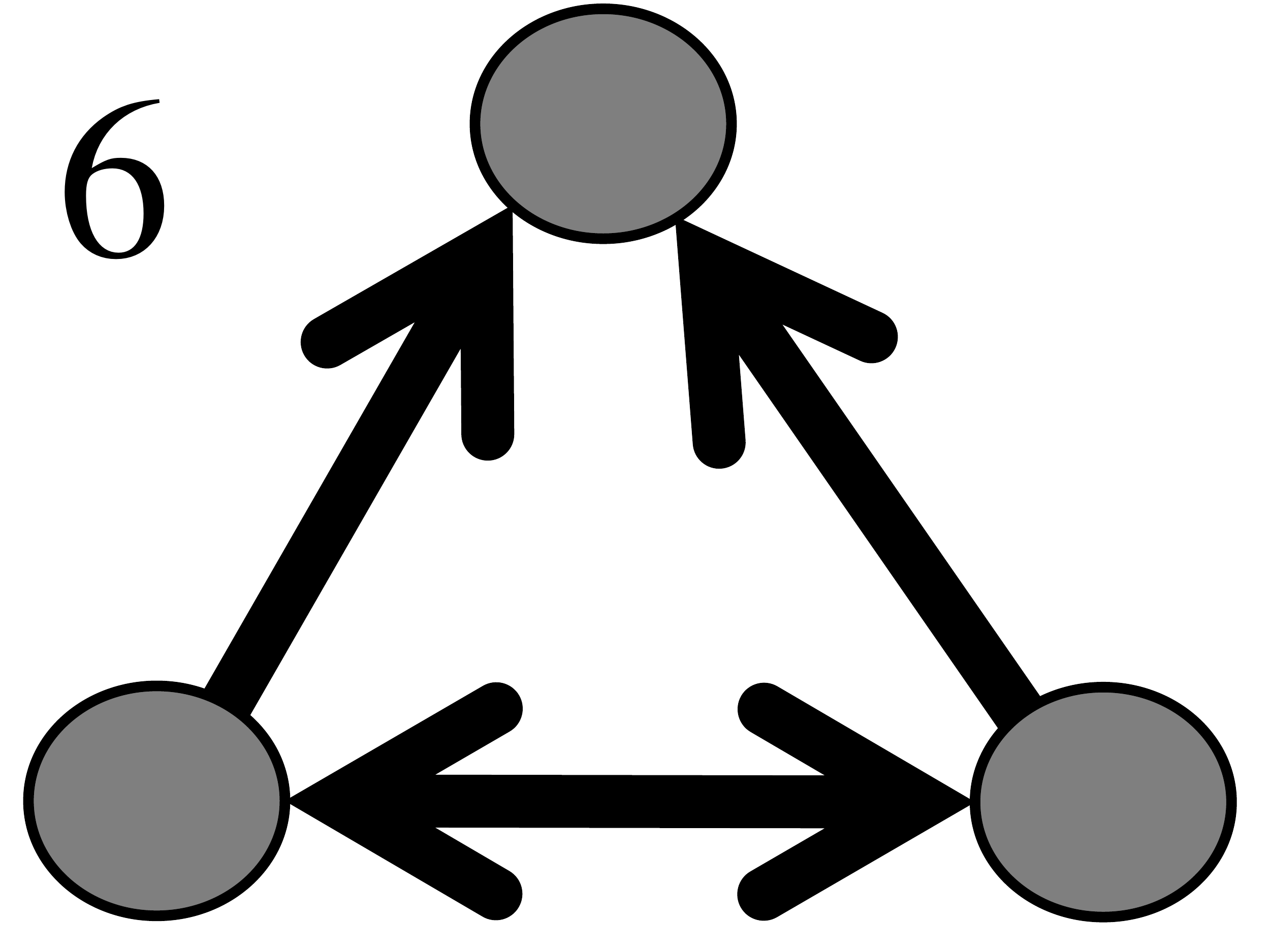}
\includegraphics[width=0.17\columnwidth]{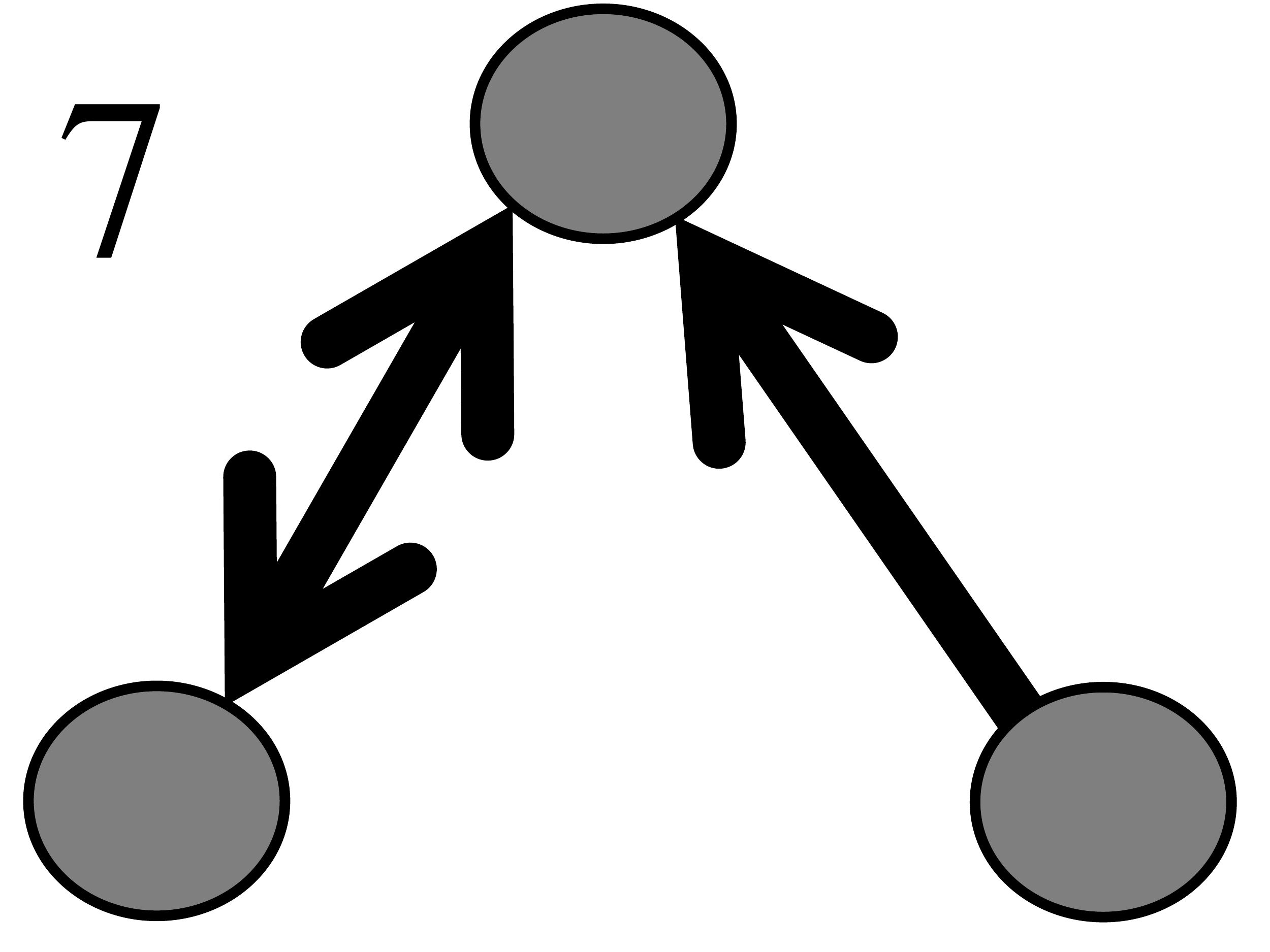}
\includegraphics[width=0.17\columnwidth]{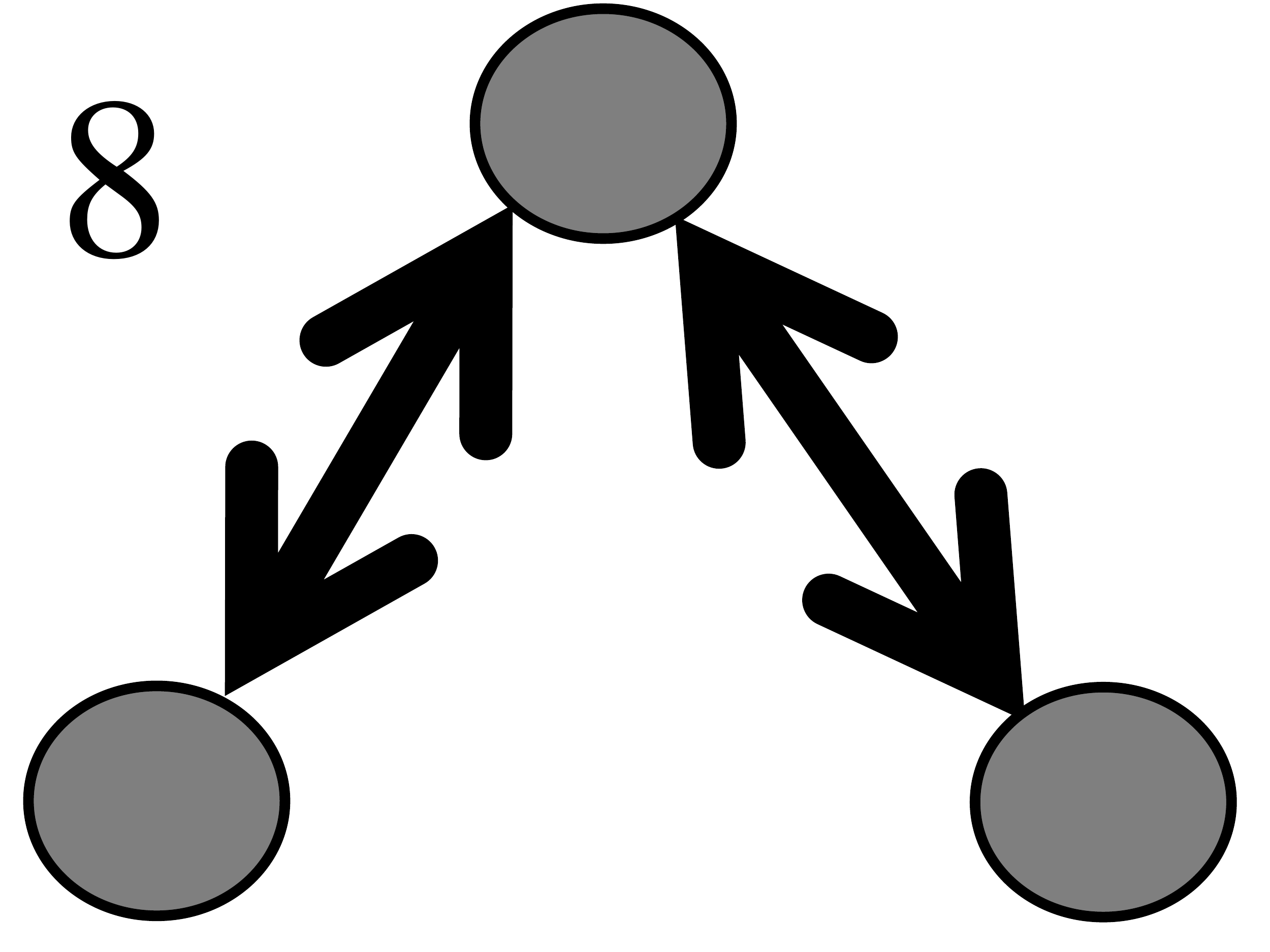}\\ \vspace{2mm}
\includegraphics[width=0.17\columnwidth]{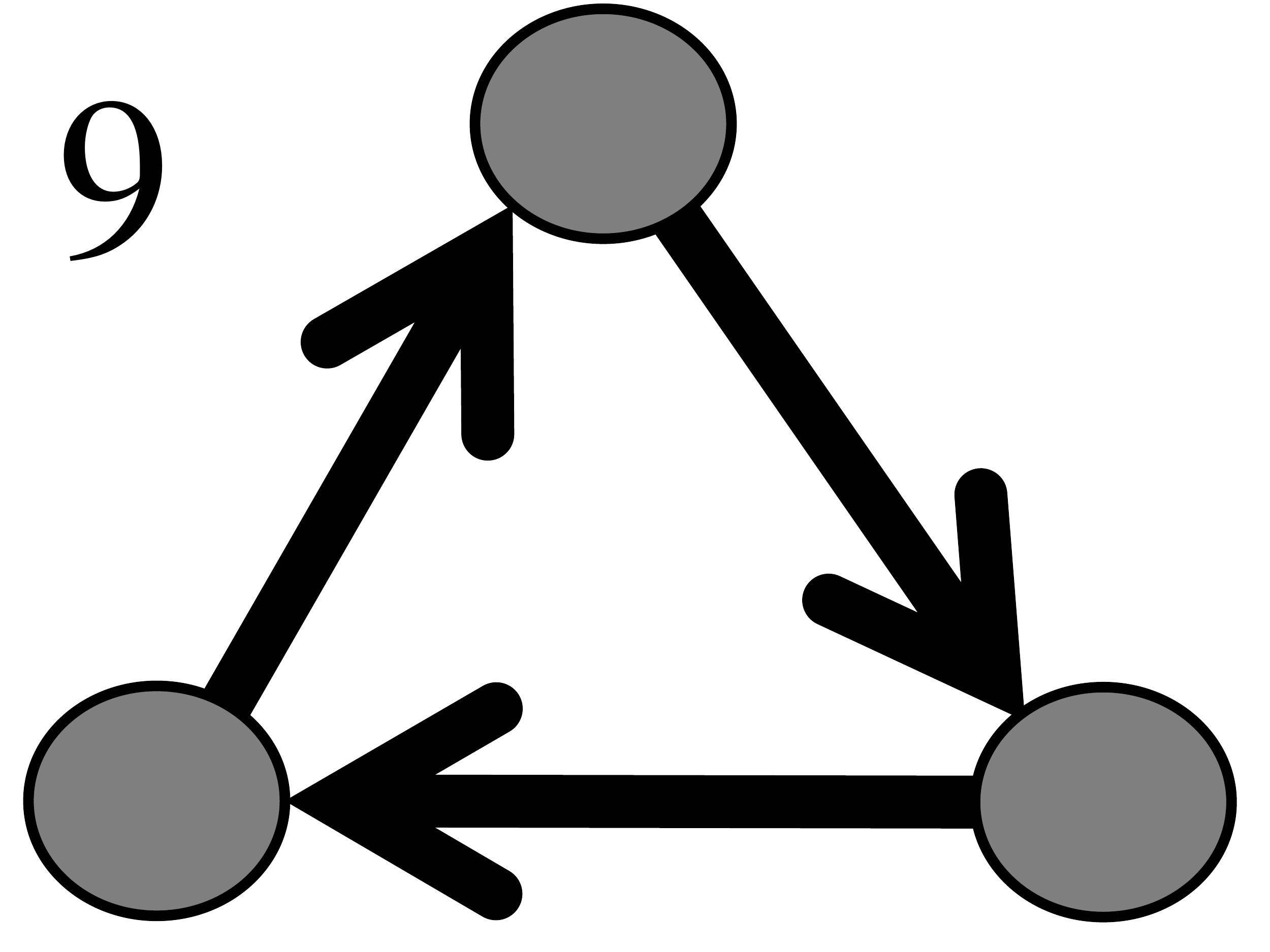}
\includegraphics[width=0.17\columnwidth]{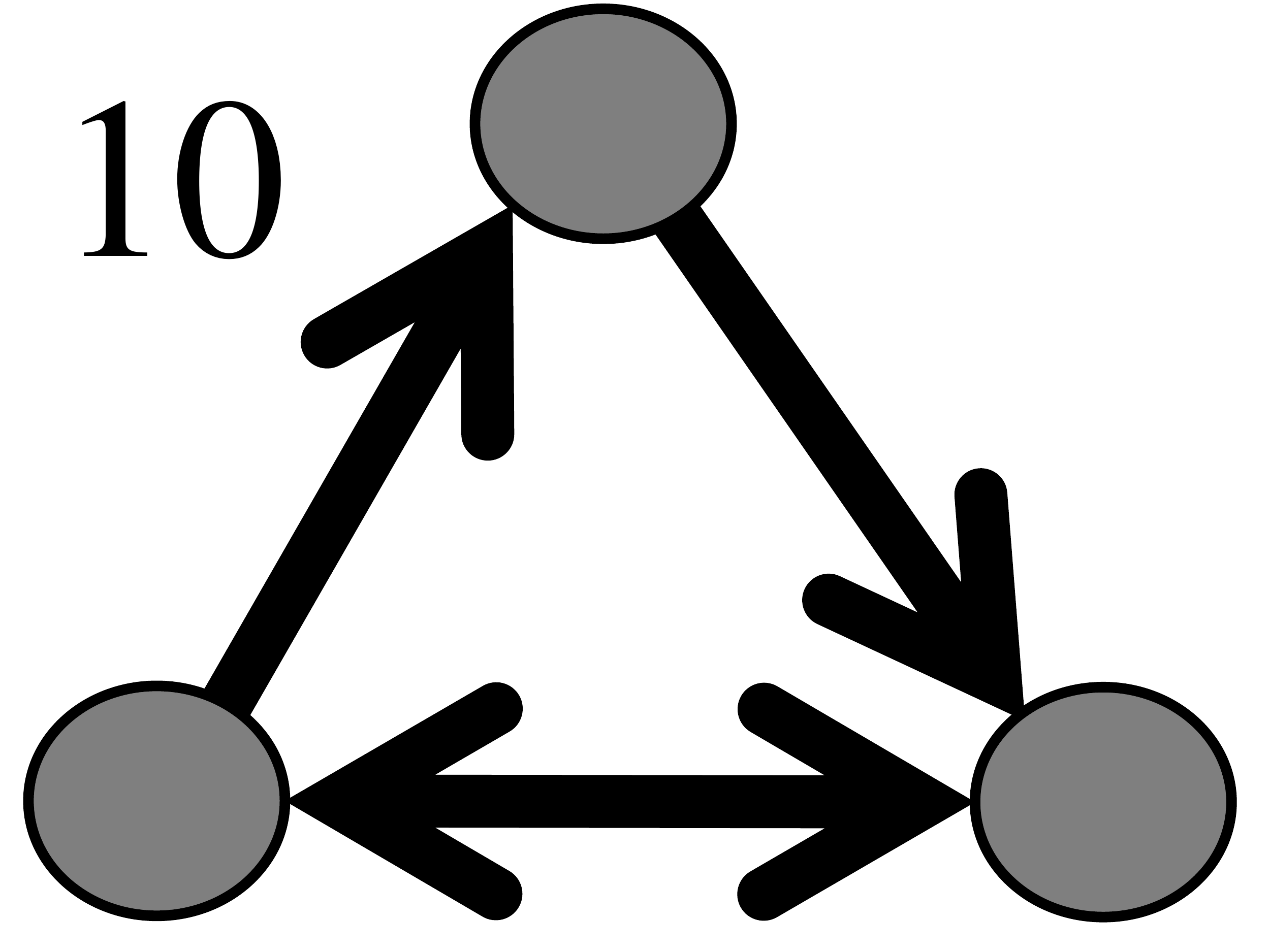}
\includegraphics[width=0.17\columnwidth]{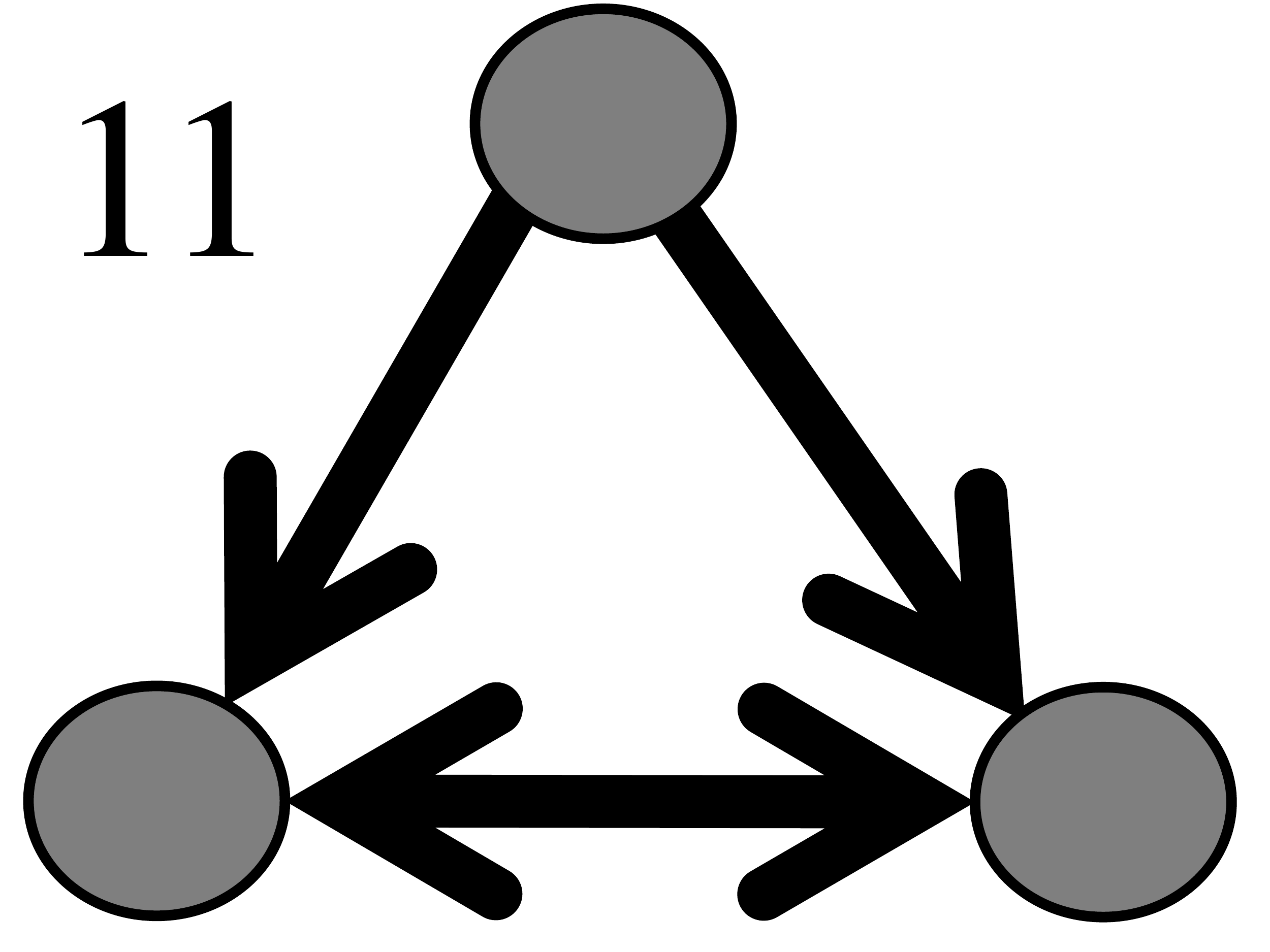}
\includegraphics[width=0.17\columnwidth]{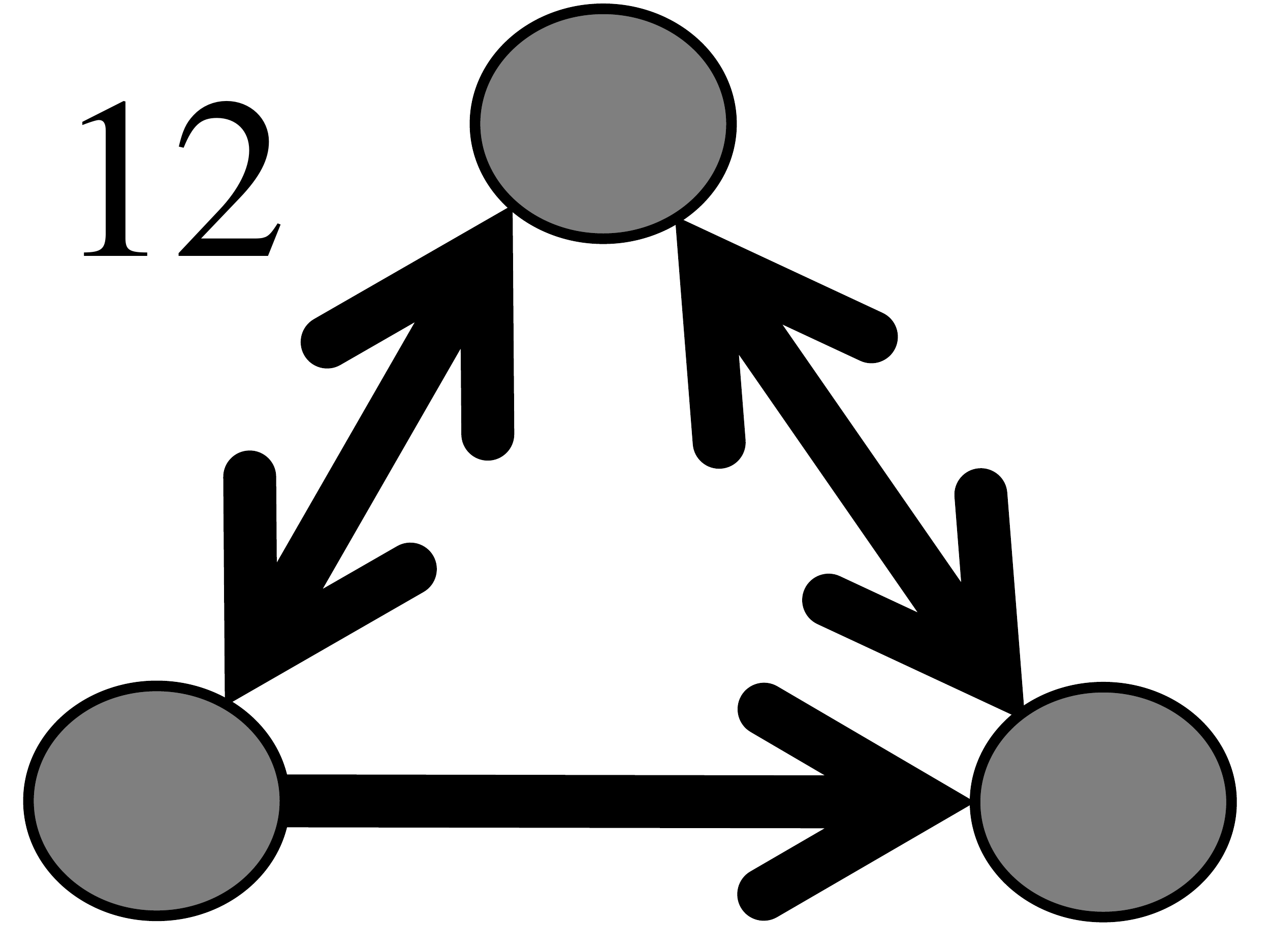}
\includegraphics[width=0.17\columnwidth]{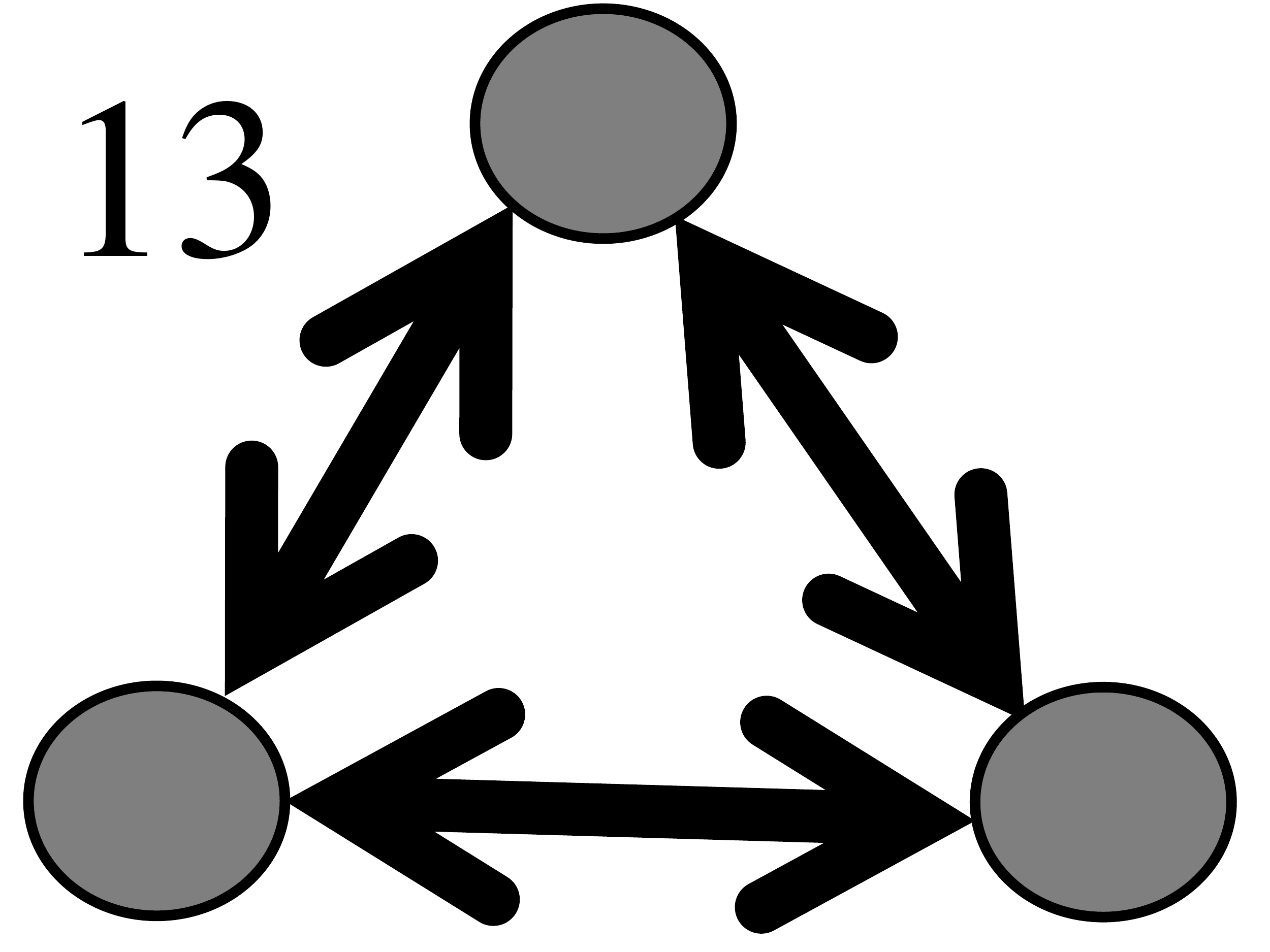}
\caption{The 13 possible non-isomorphic triangle types (cf~Definition~\ref{def:triangle}).}
\label{fig:triangle-types}
\end{figure}

As a first step, we consider the most fundamental triangle motif: weakly
connected subgraphs involving three users.  \fref{fig:triangle-types}
summarizes the thirteen non-isomorphic triangle types studied in this
paper.

For our analysis, we collected a unique data set about the user
relationships in Google$+$ OSN during a 6-week period of fast growth: in
the observed period, the network doubled in size. Our data set not only
includes information about who circles (``follows'') who, but also
meta-data about, e.g., the geographic user locations and hence the
distances of links.

To study motif changes over time, we present a snapshot-based methodology,
and make the following findings:

\begin{enumerate}
\item Slightly over $4\%$ of all triangles instances in the first snapshot
    are of a different type in the last snapshot after six weeks. How
    dynamic the triangles are depends on their type: while some motif types
    are rather stable (around 10\perc change probability), specific types
    evolve quickly in the sense that more than 30\perc of their triangles
    change into another type.  This implies that the topology structure
    changes over time.

\item The frequency distribution over different motifs is very skewed in
    all snapshots: the least frequent motif (Type~9) occurs less than
    0.001\perc, while the most frequent motif (Type~4) occurs around
    60\perc.

\item We observe a non-negligible amount of transient triangles, i.e., triangles that are not observed
if the analysis only considers triangles seen in the first or the last snapshot. This implies that to
properly capture the whole dynamics of the social network, all motifs seen during the whole lifetime of
the network must be considered.

\item Although the observed time period is characterized by a fast growth,
    more than half (50.6\perc) of all triangles that change type evolve into
    \emph{less-connected} types, i.e., many links disappear. Some even
    dissolve completely (18.6\perc). This indicates that users also prune
    their social network or change their privacy settings to hide their
    network.

\item By correlating triangles with users in- and out-degrees and user
    locations, we shed light on the semantics of different triangle types.
    Asymmetric motifs (motifs with asymmetric links only) differ in nature
    from more symmetric motifs (motifs with symmetric links): asymmetric
    edges often point to ``celebrity users'' with a high in-degree, and
    span a large geographic distance. In contrast, symmetric motifs are
    more representative of ``friendship networks'', connecting users who
    live close and who may already have met in person.
\end{enumerate}

\textbf{Organization.} The remainder of this paper is organized as follows.
Section~\ref{sec:data} introduces our methodology and describes the
collected data set.  In Section~\ref{sec:distribution}, we first study the
frequency distribution of different triangle types, and in
Section~\ref{sec:evolution} we investigate the evolution over time.  We
extend the study to include meta-data collected from user profiles in
Section~\ref{sec:profile}, and provide a discussion of our methodology and
its limitations, in Section~\ref{sec:discussion}. After reviewing related
work in Section~\ref{sec:relwork}, we conclude in
Section~\ref{sec:conclusion}.

\section{Terminology and Methodology}\label{sec:data}

In previous work~\cite{websci12google} we crawled almost the complete \gplus social network
along with publicly available profile data.  This study is based on four
snapshots $\mathcal{S}=\{S_1,\ldots,S_4\}$ from that crawl, collected over
a period of six weeks. The snapshots were taken by crawling the network on
Sep 7th (Snapshot $S_1$), Sep 20th (Snapshot $S_2$), Oct 4th (Snapshot
$S_3$), and Oct 20th (Snapshot $S_4$) in 2011. Snapshots are large: the
smallest graph has 19M nodes while the largest has 38M nodes and about 400M
edges.  Moreover, for $S_4$, we also collected publicly available profile
data, about user locations. Collection of a snapshot took almost a full day,
which in conjunction with 2 weeks in-between snapshots might lead to a bias.
Please refer to \xref{sec:discussion} for a detailed discussion.

In the following we first introduce some terminology, and then present our
data set and methodology.

\subsection{Terminology}



\noindent\textbf{Social Graph:} The \emph{social graph} consists of the set of users
(nodes) in \gplus and their relationships (directed edges) to other users expressed
through the circles.


\noindent\textbf{Node:} A user in \gplus represents a node in the graph.

\noindent\textbf{Edge:}
A (directed) \emph{edge} $A \rightarrow B$ represents the fact that user
$A$ included user $B$ in one of his circles (short: $A$ \emph{circled} $B$).
In the case that user $B$ also circled $A$, the graph contains another directed
edge $B \rightarrow A$.

\noindent\textbf{Links (asymmetric/symmetric):}
The social relation between two nodes in \gplus is called a \emph{link}. Links can
either be \emph{asymmetric} and consist of one directed edge, or they can be
\emph{symmetric}, in case the users mutually circle each other. Sometimes, we will
call a link a 2-node motif.

\noindent\textbf{Out-going, In-coming:}
The \emph{out-going} edges of a node are those directed edges that start at this node,
pointing to the members of this user's circles. The \emph{in-coming} edges of a node are
the edges that end at that node, that is, somebody else has ``circled'' the user.

\noindent\textbf{Out-degree, In-degree:} The out-degree is the number of (directed) edges
that start at a certain node. The in-degree is the number of (directed) edges that end at
a certain node.

\noindent\textbf{Neighbor:} Two nodes are neighbors if they are connected by an edge,
irrespective of the edge direction.

\noindent\textbf{Profile:} A profile is the set of personal data a user reveals about
herself. It contains the total number of in- and outgoing edges, the place(s) the user
lives, the employer, etc. This paper focuses on publicly available (``crawlable'') user
data only.

\noindent\textbf{Distance:}
The distance between nodes is calculated based on latitude and longitude as
given in a user's profile. We use the \emph{Haversine formula} to compute the great-circle distance
between two points on a sphere. Distances between nodes do not imply an edge between the nodes.

\noindent\textbf{Time Zone:}
The timezone of a node is also based on the coordinates taken from the profile. We give time
zones in absolute numbers in relation to \emph{UTC} $+0$.

We are interested in triangle motifs: the relationships between three Google$+$ users.
\begin{shaded}
\begin{definition}[Triangle]\label{def:triangle}
Three nodes $v_1,v_2,v_3$ form a \emph{triangle} if and only if the subgraph spanned by the
three nodes is \emph{weakly-connected}, i.e., the nodes are connected if the edges were undirected.
\end{definition}
\end{shaded}

Note that our triangle definition is rather general, as an edge does not have to exist between all pairs
of users in the triangle, not even a directed one. \fref{fig:triangle-types} enumerates all 13
non-isomorphic triangle types fulfilling Definition~\ref{def:triangle}. Theoretically, $x$ users can
be involved in up to $\binom{x}{3}$ triangles.

\noindent\textbf{Origin and Destination Type:} To study how the relationship between three users changes,
we compare the types of the corresponding triangles connecting them in two snapshots
$S^{(o)},S^{(d)}\in \mathcal{S}$: the \emph{origin snapshot} $S^{(o)}$ and the \emph{destination snapshot}
$S^{(d)}$. Accordingly, we call the triple relationship in $S^{(o)}$ the \emph{origin triangle} and the
relationship in $S^{(d)}$ the \emph{destination triangle}. Unless otherwise stated, we assume that
$S^{(o)}=S_1$ and $S^{(d)}=S_4$.

\noindent\textbf{Type 0:} Due to the fast growth of the network, many node
triples form a triangle according to our definition only during a subset of
the snapshots $\mathcal{S}$. To take into account triangles which only
exist for a strict subset of the snapshots $\mathcal{S}' \subsetneq
\mathcal{S}$, we introduce the notion of \emph{Type~0} triangles: an
instance of a Type~0 triangle is any relationship between three users who
were connected according to a \emph{Type~$\geq 1$} triangle in at least one
other snapshot. As we will see, Type~0 triangles do provide insight, e.g.,
regarding the triangles built by new users. Implementation wise, Type 0
triangles also help us to track the same set of triples of nodes.

\subsection{Dataset and Methodology}

Our methodology to find all triangles and what type they are works in two
steps. \emph{1)} we search all groups of three nodes that are weakly
connected, \emph{2)} we determine the type of each triple.  The number of
resulting triangles is very large. In fact we estimated the time to detect
all the triangles in a single snapshot to consume several weeks and
resulting in more disk space than we had available ($>$3\,TB). Apart from
the challenge to store such enormous amounts of data it is also hard to
process them later. So, for practical purposes, we decided to reduce the
result space. In a first step we choose four of our 16 data sets, the
first, the last, and two in between.  In the second step we removed all
nodes and their edges that do not give a meaningful location, \ie that did
not provide geo-coordinates.  Since we wanted to learn more about the user
relations in relation to their locations, e.g., the distances between
friends.

From these, we select one hundred starting nodes uniformly at random, and
then consider all triangles these nodes participate in. Within this step,
all nodes and edges that are necessary to complete the triangles are added
again to the data set. For each of the four snapshots, we start with the
same set of 100 nodes.  This method results in a set of data that is small
enough to be further processed, but big enough to preserve the graph
structure.

The sampling of the graph likely incurs a bias on the results. This is a
well known fact and reported in previous work~\cite[Section
4.1.1]{mislove-2007-socialnetworks} However similar work relies on similar
problem space reduction techniques as we do~\cite{Cha-Flickr,
mislove-2007-socialnetworks}. For a discussion of the bias of our
methodology we refer the reader to Section~\ref{sec:discussion}.

\begin{table}[t]
\centering
\caption{Graphs}\label{tab:graphs}
\tabcolsep3mm
\begin{tabular}{llrrr}
\toprule
\multicolumn{2}{l}{\btmr{Graph (Date in 2011)}}	
				& \multicolumn{3}{c}{Number of (in Mio.)} \\
\cmidrule{3-5}
	&			& Nodes & Edges & Triangles \\
\midrule
Sep 7	& full 			&  19.6 & 278.3 &   -- \\
Sep 7	& locations only 	&   3.3 &  43.8 &   -- \\
Sep 7	& triangle-graph 	&   1.6 &  29.3 &  4.9 \\
\midrule
Sep 20	& full 			&  20.7 & 294.4 &   -- \\
Sep 20	& locations only 	&   3.5 &  46.4 &   -- \\
Sep 20	& triangle-graph 	&   1.6 &  31.2 &  6.3 \\
\midrule
Oct 4	& full 			&  36.2 & 388.3 &   -- \\
Oct 4	& locations only 	&   8.8 &  64.6 &   -- \\
Oct 4	& triangle-graph 	&   2.7 &  41.0 &  8.1 \\
\midrule
Oct 20	& full 			&  38.6 & 476.9 &   -- \\
Oct 20	& locations only 	&   9.6 &  83.7 &   -- \\
Oct 20	& triangle-graph    	&   3.4 &  54.5 &  8.9 \\
\bottomrule
\end{tabular}
\end{table}

\tref{tab:graphs} gives an overview of the graphs considered in this
paper. Here, ``full'' refers to the graph representing the entire snapshot.
The step in between, stripping the graph of all nodes without location
information, is referred to as ``location only'' triangles: graphs where
nodes without location information in the profile are ignored (together
with their incident edges).  The ``triangle-graph rand'' category contains
the triangles resulting from choosing one hundred starting nodes at random.


\subsection{Interpretation of Motifs}

The \gplus  ``social search OSN'' occupies a peculiar position between a friendship network
and a news aggregator network. In this paper, we will sometimes
interpret the semantics of different motifs accordingly (see also the
discussion in Section~\ref{sec:discussion}): We argue that more symmetric
motifs (where users are mutually connected) are an indication of a friendship
relationship, while more asymmetric relationships suggest that a user follows
someone he or she has not yet met in person.


\begin{table}
\centering
\caption{Superstars}\label{tab:superstars}
\tabcolsep2mm
\begin{tabular}{lcccc}
\toprule
\btmr{Name}
		&  \btmr{Rank}	
			& \btmr{\parbox{1cm}{\centering Followers in full}} 	
					& \multicolumn{2}{c}{\perc of Followers in full} \\
\cmidrule{4-5}
		& 	&   		& location only & triangle-graph\\
\midrule
Britney Spears  &     1 & 443854 	& 62.11\perc 	& 59.29\perc \\
Mark Zuckerberg &     2 & 510132 	& 45.74\perc 	& 44.99\perc \\
Paris Hilton    &     3 & 336174 	& 66.07\perc 	& 64.72\perc \\
Sergey Brin     &     4 & 351943 	& 58.57\perc 	& 57.80\perc \\
Jessi June      &     5 & 275122 	& 64.29\perc 	& 63.31\perc \\
Vic Gundotra    &     6 & 277713 	& 60.41\perc 	& 60.09\perc \\
Mark Cuban      &     7 & 219765 	& 61.66\perc 	& 59.81\perc \\
Thomas Hawk     &     8 & 204357 	& 61.84\perc 	& 61.49\perc \\
Trey Ratcliff   &     9 & 205413 	& 58.27\perc 	& 58.11\perc \\
Pitbull         &    10 & 187610 	& 62.23\perc 	& 59.61\perc \\
\bottomrule
\end{tabular}
\end{table}

We observe several indicators that lead to this interpretation.
First, we observe that asymmetric links often point to users with a high
in-degree. For example \tref{tab:superstars} shows the top-10
in-degree users. Depending on the connection between the followers
the resulting triangle type is 4, 5, or 6.  Second, as we observed
earlier~\cite{websci12google} the geographic distances spanned by
asymmetric links are generally larger than for symmetric links.


\section{Motif Distribution: A First Look}\label{sec:distribution}


\begin{table*}[t]
\centering
\caption{Relative frequency of triangle types across datasets}
\label{tab:tria-perc}
\tabcolsep2mm
\begin{tabular}{lccccccccccccc}
\toprule
\btmr{Data} 	& \multicolumn{13}{c}{Frequency of triangle type (in\perc)} \\
\cmidrule{2-14}
		&  1 \tria  & 2 \trib & 3 \tric & 4 \trid & 5 \trie & 6
		\trif & 7 \trig & 8 \trih & 9 \trii & 10 \trij & 11 \trik &
		12 \tril & 13 \trim \\
\midrule
Sep 7 	& 20.86 & 6.82 & 5.23 & 55.75 & 3.18 & 0.40 & 4.74 & 1.43 & 0.00002 & 0.08 & 1.16 & 0.28 & 0.03 \\
Sep 20 	& 17.37 & 5.56 & 4.46 & 62.97 & 2.74 & 0.40 & 3.87 & 1.34 & 0.00006 & 0.07 & 0.96 & 0.26 & 0.03  \\
Oct 4	& 12.91 & 6.37 & 5.12 & 63.79 & 2.86 & 0.87 & 4.29 & 2.22 & 0.00005 & 0.08 & 0.90 & 0.46 & 0.14  \\
Oct 20	& 15.12 & 5.22 & 5.49 & 66.48 & 2.13 & 0.66 & 3.09 & 0.87 & 0.00024 & 0.05 & 0.68 & 0.20 & 0.02 \\
\bottomrule
\end{tabular}
\end{table*}

We first study the frequency distribution of different motifs in Google$+$
snapshots.  As one might expect, the distribution is quite skewed, see
\tref{tab:tria-perc}: Type 4 is by far the most frequent motif, with
over 50 percent of all triangles of this type.  According to our motif
interpretation, this means that a large fraction of users follows other
users. The peculiar Type~9 is the least frequent motif: indeed, it
describes a situation where three users are connected in a circular manner,
where the circled user does not include the original user. Type 13
triangles (three mutually connected users) are quite rare: this can be seen
as a further indication that Google$+$ is used for more than just keeping
up with friends.

Another take-away from \tref{tab:tria-perc} is that triangles are
typically sparse, i.e., triangles with the minimum of only two directed
edges constitute the vast majority of all the triangles.  Among the motifs
where all three nodes are directly linked, Type~6 is quite frequent: it
describes a situation where two mutually connected users (\eg friends)
follow the same third user. Much more frequent however is Type~5, where
there is only an asymmetric link between two ``followers''. Among the
triangle motifs with at least one symmetric link, Type~3 and Type~7 occur
most often---again, essentially the motifs with the least edges fulfilling
the symmetric criteria.

\section{Motif Dynamics}\label{sec:evolution}

Active social networks are in a constant flux: new users join the system,
existing users update their relationships over time, and others leave. This
is particularly true for the young Google$+$ network: The observed period
features a significant growth (the user base and number of triangles more
than doubled), but many users also remove links during this period of
growth: the fraction of triangles evolving into a sparser motif is quite
large.

This section presents our first insights on how triangles change their type
between crawled snapshots.  Unless otherwise stated, we will focus on the
triangles sampled for the snapshot $S_1$ (Sep 7), and compare a given
triangle's type in $S_1$ (the \emph{origin type} or origin triangle) with
its type in $S_4$ (the \emph{destination type} or destination triangle).
We first consider all triangle transitions including those from and to
Type~0, in order to also understand the creation and destruction process of
triangles. Then we focus on triangle to triangle transitions.

\subsection{Triangle Transitions Including Type~0}

Due to the fast growth of the user base, many triangles only emerge after
snapshot $S_1$ or disappear before $S_4$. Accordingly, in this section, we
consider Type~0 triangles: user triples who formed a valid triangle for one
or more snapshots in $\mathcal{S}$.

In our 6-week observation period we observe the majority of triangles
changing type. We observe for 61.57\% of node triples that at lease once
form a triangle a change in their interconnection. This implies that a
significant amount of transient dynamics within a social network may be
lost when only the triangles present in the original or final snapshot are
considered.

\begin{table}[t]
\centering
\caption{Triangle type transition probability (in\perc)}
\label{tab:change-pertype}
\tabcolsep1mm
\begin{tabular}{cccccccccccccc}
\toprule
0 & 1 & 2 & 3 & 4 & 5 & 6 & 7 & 8 & 9 & 10 & 11 & 12 & 13 \\
\midrule
100.0 & 32.2 & 16.5 & 20.5 & 20.0 & 9.8 & 10.9 & 21.2 & 27.8 & 100.0 & 18.4
& 12.2 & 9.4 & 11.5 \\
\bottomrule
\end{tabular}
\end{table}

\tref{tab:change-pertype} shows the probability that a certain type
changes. Note, that all Type~0 triangle need to change, per definition.
Also note that there is only one Type~9 triangle in the first snapshot. For
the other types we observe quite varying change probabilities. This
indicates that some types are more stable than others. Types 5, 6, 11, 12,
and 13 are the most stable ones with change rates around 10\perc. Types 1
and 8 expose the highest change rates around 30\perc. Overall however most
triangle do not change at all.

\begin{figure}[t]
\centering
\includegraphics[width=1.0\columnwidth,angle=0]{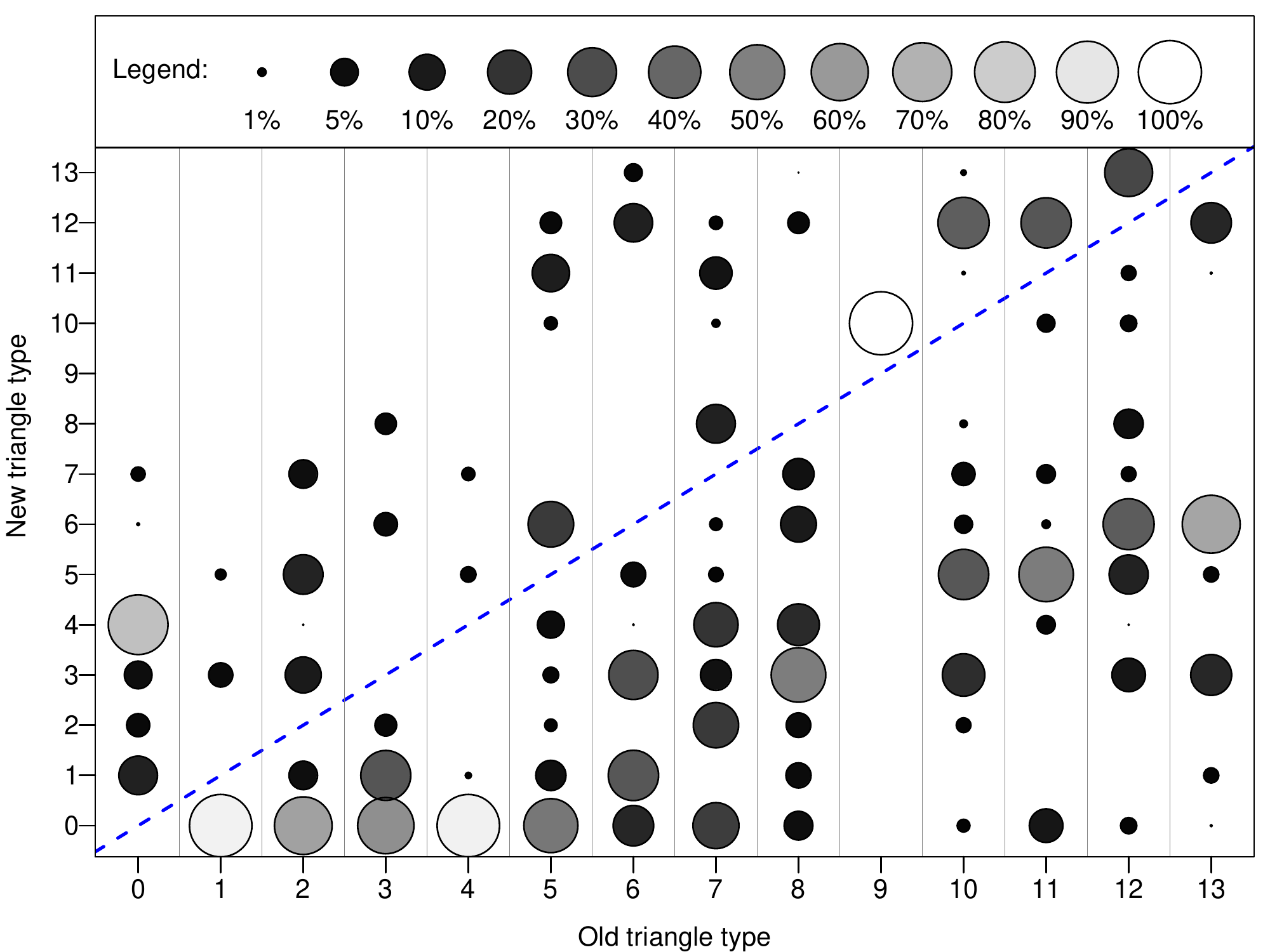}
\caption{Triangle transition destination types per triangle type: Each
column of circles adds up to 100\perc and indicates the distribution of
destination types per triangle type.}
\label{fig:trans-pertype-dest}
\end{figure}

\begin{figure}[t]
\centering
\includegraphics[width=1.0\columnwidth,angle=0]{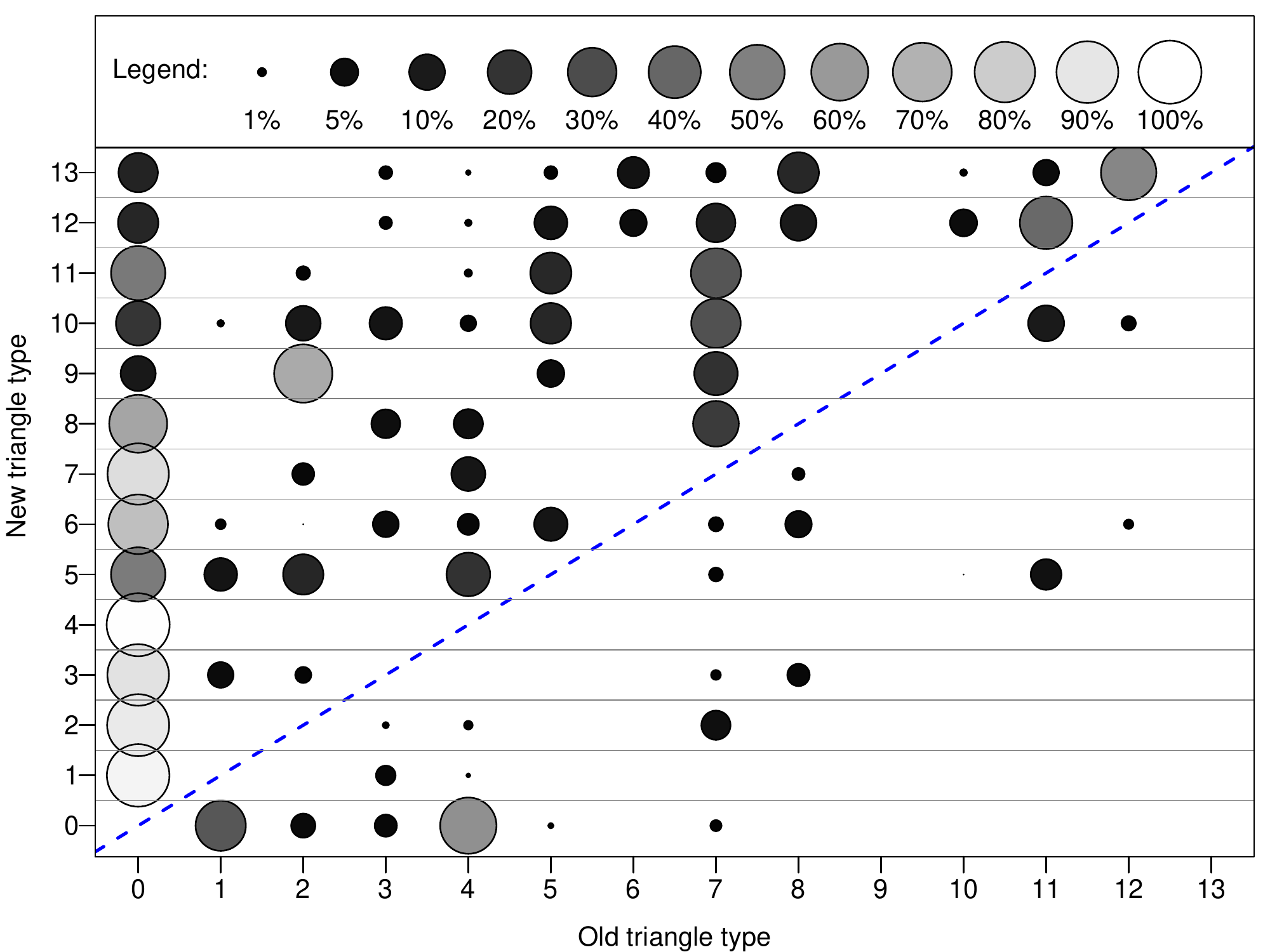}
\caption{Triangle transition origin types per triangle type: Each
row of circles adds up to 100\perc and indicates the distribution of
origin types per triangle type.}
\label{fig:trans-pertype-orig}
\end{figure}

\fref{fig:trans-pertype-orig} shows the distribution of destination types
triangles transition into per origin triangle type. So for example the plot
shows that in relation to all transitions with an origin Type 4 ($x=4$),
94.6\perc turn into a Type 0 triangle ($y=0$), and Type~5, 7, and 1 are
the next most likely destination with 1.8\perc, 1.5\perc, and 0.8\perc.
Note that we only plot a circle when the transition probability is higher than 0.5\perc.

Transitions below the diagonal in \fref{fig:trans-pertype-orig}, indicate a
degeneration of the triangle, typically involving the loss of an edge.
Looking at the figure it is obvious that more triangles degenerate (below
the diagonal) as compared to those that strengthen their relation (above
the diagonal). It is also apparent that Type~0 is the most likely
destination for origin types 1 through 5, \ie those triangles disintegrate.
For the better connected triangle types 10 through 13, however types 5 and
6 are the most likely destination types.

In \fref{fig:trans-pertype-dest} we now look at the origin type instead of
the destination type. Here, for almost every destination type (y-axis),
except types 9, 10, 12, and 13 we see that Type~0 is the most likely origin
type. Yet, interesting enough around 14\perc of types 12 and 13 evolve
directly from Type~0. Obviously, our snapshots are two far apart to capture
the full dynamics, but none the less this indicates that Type~13 triangles
can be created in short time frames of around 2-weeks.

This general trend towards sparser motifs may have multiple possible
reasons, including changes in privacy settings (\ie thereby hiding links
from our crawling method), people dropping out from Google$+$, or active
pruning of users from \emph{Circles} due to changing interests. We expect
that most of the vanishing edges are due to changes in privacy settings,
however in some occasions it is reasonable to expect that users stop to
follow another user for various reasons, including but not limited to a too
high posting frequency, to boring content shared by the followed, or actual
change in off-line relationships. However, our data does not provide direct
evidence to argue about which of the reasons is more prevalent.


%
%
%


\subsection{Triangle-to-Triangle Transitions}


Given the large impact of transitions involving Type~0, we now turn to
triangles that already existed in $S_1$, and whose user triple still formed
a connected triangle in $S_4$.

\begin{figure}[t]
\centering
\includegraphics[width=1.0\columnwidth,angle=0]{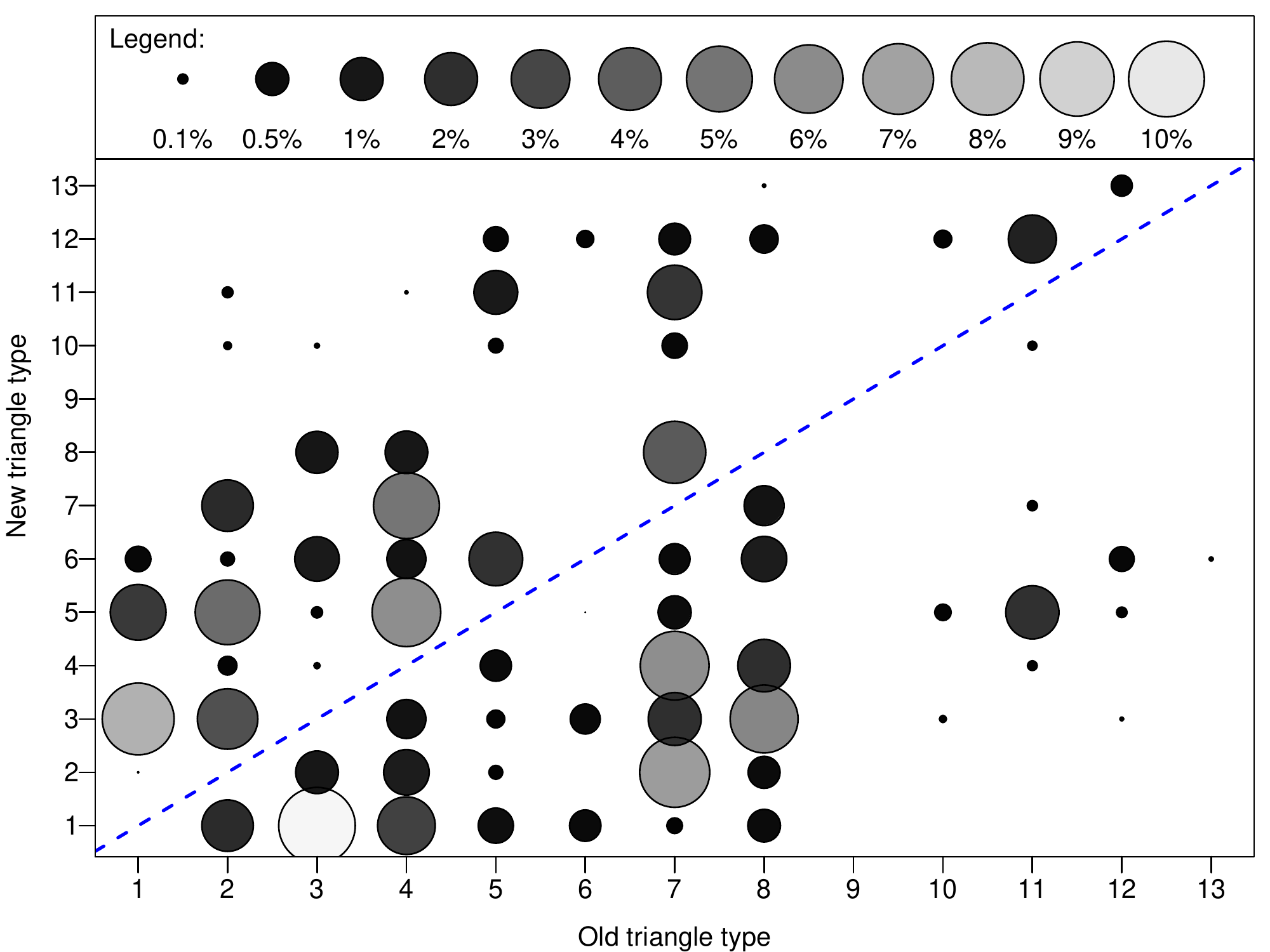}
\caption{Transition of triangle types without Type~0. We observe 4.12\perc
of triangles transition to a new type. Note this plot does not show
per-type frequencies. The biggest circle represents
10.6\perc of all triangle to triangle transitions. See also table
\tref{tab:trans-wo-type0}.}
\label{fig:transbubble}
\end{figure}

\tref{tab:trans-wo-type0} gives an overview of
frequencies of all transitions (\ie not per type), see also
\fref{fig:transbubble} for a graphical representation.  Without
Type~0 triangles, 4.12\perc of the triangles in the first snapshot $S_1$
changed to a different type in the last snapshot $S_4$.

\begin{table*}[t]
  \centering
  \caption{All triangle transitions between first and last snapshot
  (without Type~0). See also \fref{fig:transbubble}. Overall, 4.12\%
  of the triangles changed their type.}\label{tab:trans-wo-type0}
\tabcolsep2mm
\begin{tabular}{cccccccccccccc}
\toprule
   &      1 &      2 &      3 &      4 &      5 &      6 &      7 &      8 &      9 &     10 &     11 &     12 &     13 \\
\midrule
 1 &    --- &  0.055 &  7.653 &  0.006 &  2.473 &  0.297 &  0.000 &  0.023 &  0.000 &  0.008 &  0.027 &  0.009 &  0.001 \\
 2 &  1.835 &    --- &  3.442 &  0.184 &  4.626 &  0.131 &  1.827 &  0.032 &  0.009 &  0.085 &  0.106 &  0.010 &  0.000 \\
 3 & 10.599 &  0.986 &    --- &  0.077 &  0.109 &  1.122 &  0.028 &  0.942 &  0.000 &  0.071 &  0.000 &  0.050 &  0.006 \\
 4 &  2.824 &  1.204 &  0.762 &    --- &  6.104 &  0.760 &  5.039 &  0.981 &  0.000 &  0.017 &  0.063 &  0.031 &  0.003 \\
 5 &  0.588 &  0.132 &  0.173 &  0.443 &    --- &  2.131 &  0.001 &  0.000 &  0.001 &  0.138 &  1.066 &  0.280 &  0.006 \\
 6 &  0.450 &  0.005 &  0.407 &  0.007 &  0.052 &    --- &  0.001 &  0.001 &  0.000 &  0.001 &  0.000 &  0.164 &  0.029 \\
 7 &  0.149 &  6.740 &  2.027 &  6.140 &  0.505 &  0.428 &    --- &  3.880 &  0.002 &  0.291 &  2.253 &  0.456 &  0.010 \\
 8 &  0.491 &  0.468 &  5.802 &  1.969 &  0.049 &  1.188 &  0.813 &    --- &  0.000 &  0.003 &  0.004 &  0.357 &  0.063 \\
 9 &  0.000 &  0.000 &  0.000 &  0.000 &  0.000 &  0.000 &  0.000 &  0.000 &    --- &  0.001 &  0.000 &  0.000 &  0.000 \\
10 &  0.001 &  0.008 &  0.082 &  0.001 &  0.158 &  0.010 &  0.016 &  0.004 &  0.000 &    --- &  0.003 &  0.170 &  0.004 \\
11 &  0.008 &  0.005 &  0.020 &  0.098 &  2.066 &  0.043 &  0.101 &  0.002 &  0.000 &  0.093 &    --- &  1.442 &  0.018 \\
12 &  0.003 &  0.003 &  0.065 &  0.004 &  0.104 &  0.283 &  0.014 &  0.045 &  0.000 &  0.015 &  0.014 &    --- &  0.217 \\
13 &  0.002 &  0.000 &  0.016 &  0.000 &  0.002 &  0.067 &  0.000 &  0.000 &  0.000 &  0.000 &  0.001 &  0.015 &    --- \\
\bottomrule
\end{tabular}
\end{table*}

We extract several take-away's from
\tref{tab:trans-wo-type0} and
\fref{fig:transbubble}.  First, we observe that the transition
probabilities are rather asymmetric. This indicates that the distribution
of motif frequencies, and hence the character of the topology, is changing
over time. Another interesting observation is that
the rate of change, as well as the most likely predecessor and successor
types, depend on the triangle type. Taking into account the relative
frequencies of the corresponding types (see \tref{tab:tria-perc}), we
can compute the relative change frequency per type, see
\tref{tbl:summary}; the table also includes the most frequent
predecessor and successor type for each triangle. The table shows that over
25$\%$ of all Type~8 triangles change between $S_1$ and $S_4$. Type~10 and
Type~7 also change very frequently (more than $17 \%$). In contrast,
Types~1,~2, and~5 are quite stable (change rate around 5$\%$). Maybe
surprisingly, the change frequency of Type~9 is low; however, due to the
small absolute number of triangles of this type, we believe that
statistical significance is insufficient. \tref{tbl:summary} also
shows that for some triangle types, namely Types~1,~3,~7, and~8, the
predecessor and the successor types are the same.

\begin{table}[h!]
  \centering
  \begin{tabular}{| c | c | c | c | }
 \hline
  type & pred & succ & freq
   \\ \hline
   \begin{minipage}[c][1cm][c]{.12\textwidth} \center
      \includegraphics[height=7mm]{tri-1doris}
    \end{minipage}
    &
   \begin{minipage}[c][1cm][c]{.12\textwidth} \center
      \includegraphics[height=7mm]{tri-3doris}
    \end{minipage}
    &
  \begin{minipage}[c][1cm][c]{.12\textwidth} \center
      \includegraphics[height=7mm]{tri-3doris}
    \end{minipage}
    &
    2.4 $\%$
   \\ \hline
   \begin{minipage}[c][1cm][c]{.12\textwidth} \center
      \includegraphics[height=7mm]{tri-2doris}
    \end{minipage}
    &
   \begin{minipage}[c][1cm][c]{.12\textwidth} \center
      \includegraphics[height=7mm]{tri-7doris}
    \end{minipage}
    &
  \begin{minipage}[c][1cm][c]{.12\textwidth} \center
      \includegraphics[height=7mm]{tri-5doris}
    \end{minipage}
    &
6.7\%
   \\ \hline
   \begin{minipage}[c][1cm][c]{.12\textwidth} \center
      \includegraphics[height=7mm]{tri-3doris}
    \end{minipage}
    &
   \begin{minipage}[c][1cm][c]{.12\textwidth} \center
      \includegraphics[height=7mm]{tri-1doris}
    \end{minipage}
    &
  \begin{minipage}[c][1cm][c]{.12\textwidth} \center
      \includegraphics[height=7mm]{tri-1doris}
    \end{minipage}
&
10.1\%
   \\ \hline
   \begin{minipage}[c][1cm][c]{.12\textwidth} \center
      \includegraphics[height=7mm]{tri-4doris}
    \end{minipage}
    &
   \begin{minipage}[c][1cm][c]{.12\textwidth} \center
      \includegraphics[height=7mm]{tri-7doris}
    \end{minipage}
    &
  \begin{minipage}[c][1cm][c]{.12\textwidth} \center
      \includegraphics[height=7mm]{tri-5doris}
    \end{minipage}
&
1.3$\%$
   \\ \hline
   \begin{minipage}[c][1cm][c]{.12\textwidth} \center
      \includegraphics[height=7mm]{tri-5doris}
    \end{minipage}
    &
   \begin{minipage}[c][1cm][c]{.12\textwidth} \center
      \includegraphics[height=7mm]{tri-4doris}
    \end{minipage}
    &
  \begin{minipage}[c][1cm][c]{.12\textwidth} \center
      \includegraphics[height=7mm]{tri-6doris}
    \end{minipage}
&
5.4$\%$
   \\ \hline
   \begin{minipage}[c][1cm][c]{.12\textwidth} \center
      \includegraphics[height=7mm]{tri-6doris}
    \end{minipage}
    &
   \begin{minipage}[c][1cm][c]{.12\textwidth} \center
      \includegraphics[height=7mm]{tri-5doris}
    \end{minipage}
    &
  \begin{minipage}[c][1cm][c]{.12\textwidth} \center
      \includegraphics[height=7mm]{tri-1doris}
    \end{minipage}
&
9.4$\%$
   \\ \hline
   \begin{minipage}[c][1cm][c]{.12\textwidth} \center
      \includegraphics[height=7mm]{tri-7doris}
    \end{minipage}
    &
   \begin{minipage}[c][1cm][c]{.12\textwidth} \center
      \includegraphics[height=7mm]{tri-4doris}
    \end{minipage}
    &
  \begin{minipage}[c][1cm][c]{.12\textwidth} \center
      \includegraphics[height=7mm]{tri-4doris}
    \end{minipage}
&
17 \%
   \\ \hline
   \begin{minipage}[c][1cm][c]{.12\textwidth} \center
      \includegraphics[height=7mm]{tri-8doris}
    \end{minipage}
    &
   \begin{minipage}[c][1cm][c]{.12\textwidth} \center
      \includegraphics[height=7mm]{tri-7doris}
    \end{minipage}
    &
  \begin{minipage}[c][1cm][c]{.12\textwidth} \center
      \includegraphics[height=7mm]{tri-3doris}
    \end{minipage}
&
26.6$\%$
   \\ \hline
   \begin{minipage}[c][1cm][c]{.12\textwidth} \center
      \includegraphics[height=7mm]{tri-9doris}
    \end{minipage}
    &
   \begin{minipage}[c][1cm][c]{.12\textwidth} \center
      \includegraphics[height=7mm]{tri-1doris}
    \end{minipage}
    &
  \begin{minipage}[c][1cm][c]{.12\textwidth} \center
      \includegraphics[height=7mm]{tri-10doris}
    \end{minipage}
& 0.0 $\%$
   \\ \hline
   \begin{minipage}[c][1cm][c]{.12\textwidth} \center
      \includegraphics[height=7mm]{tri-10doris}
    \end{minipage}
    &
   \begin{minipage}[c][1cm][c]{.12\textwidth} \center
      \includegraphics[height=7mm]{tri-7doris}
    \end{minipage}
    &
  \begin{minipage}[c][1cm][c]{.12\textwidth} \center
      \includegraphics[height=7mm]{tri-12doris}
    \end{minipage}
& 18.2$\%$
   \\ \hline
   \begin{minipage}[c][1cm][c]{.12\textwidth} \center
      \includegraphics[height=7mm]{tri-11doris}
    \end{minipage}
    &
   \begin{minipage}[c][1cm][c]{.12\textwidth} \center
      \includegraphics[height=7mm]{tri-7doris}
    \end{minipage}
    &
  \begin{minipage}[c][1cm][c]{.12\textwidth} \center
      \includegraphics[height=7mm]{tri-5doris}
    \end{minipage}
& 11.3$\%$
   \\ \hline
   \begin{minipage}[c][1cm][c]{.12\textwidth} \center
      \includegraphics[height=7mm]{tri-12doris}
    \end{minipage}
    &
   \begin{minipage}[c][1cm][c]{.12\textwidth} \center
      \includegraphics[height=7mm]{tri-11doris}
    \end{minipage}
    &
  \begin{minipage}[c][1cm][c]{.12\textwidth} \center
      \includegraphics[height=7mm]{tri-6doris}
    \end{minipage}
& 9.2$\%$
   \\ \hline
   \begin{minipage}[c][1cm][c]{.12\textwidth} \center
      \includegraphics[height=7mm]{tri-13doris}
    \end{minipage}
    &
   \begin{minipage}[c][1cm][c]{.12\textwidth} \center
      \includegraphics[height=7mm]{tri-12doris}
    \end{minipage}
    &
  \begin{minipage}[c][1cm][c]{.12\textwidth} \center
      \includegraphics[height=7mm]{tri-6doris}
    \end{minipage}
& 11.4$\%$
\\ \hline
   \end{tabular}
  \caption{Overview of most frequent origin and destination types as well as change frequency.}\label{tbl:summary}
\end{table}


%
%

\section{Motif Contexts}\label{sec:profile}

Triangle types differ by much more than just their statistical frequencies.
While a semantic characterization of different directed triangles is beyond the scope of this
paper (see Section~\ref{sec:discussion} for a short discussion), this section gives some
insights into the \emph{context} in which a triangle usually appears.

Concretely, we investigate the correlation between triangle type and user degrees.
While naturally, many Type~4 triangles are likely to occur together, in the sense that the circled
user is a celebrity with a high in-degree (describing other Type~4 triangles),
we will see that the triangle degrees also correlate with the change rate of the corresponding triangles.

Subsequently, we will take into account additional, publicly available profile data,
and show that different triangle types also differ in the distances spanned by their edges.

\subsection{Relationship Between Types and Degrees}

\begin{figure*}[t]
\centering
\includegraphics[width=.32\linewidth,angle=0]{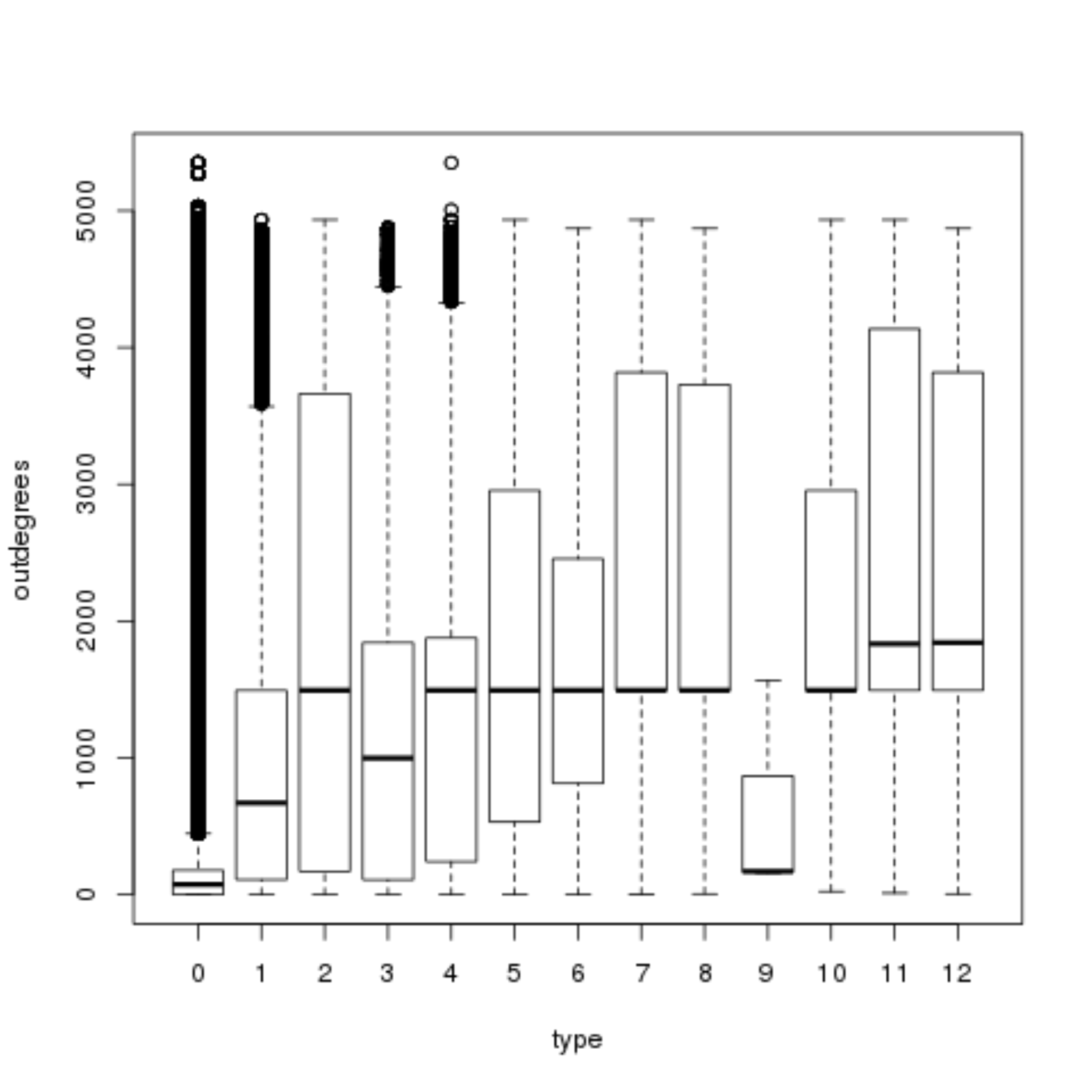}
\includegraphics[width=.32\linewidth,angle=0]{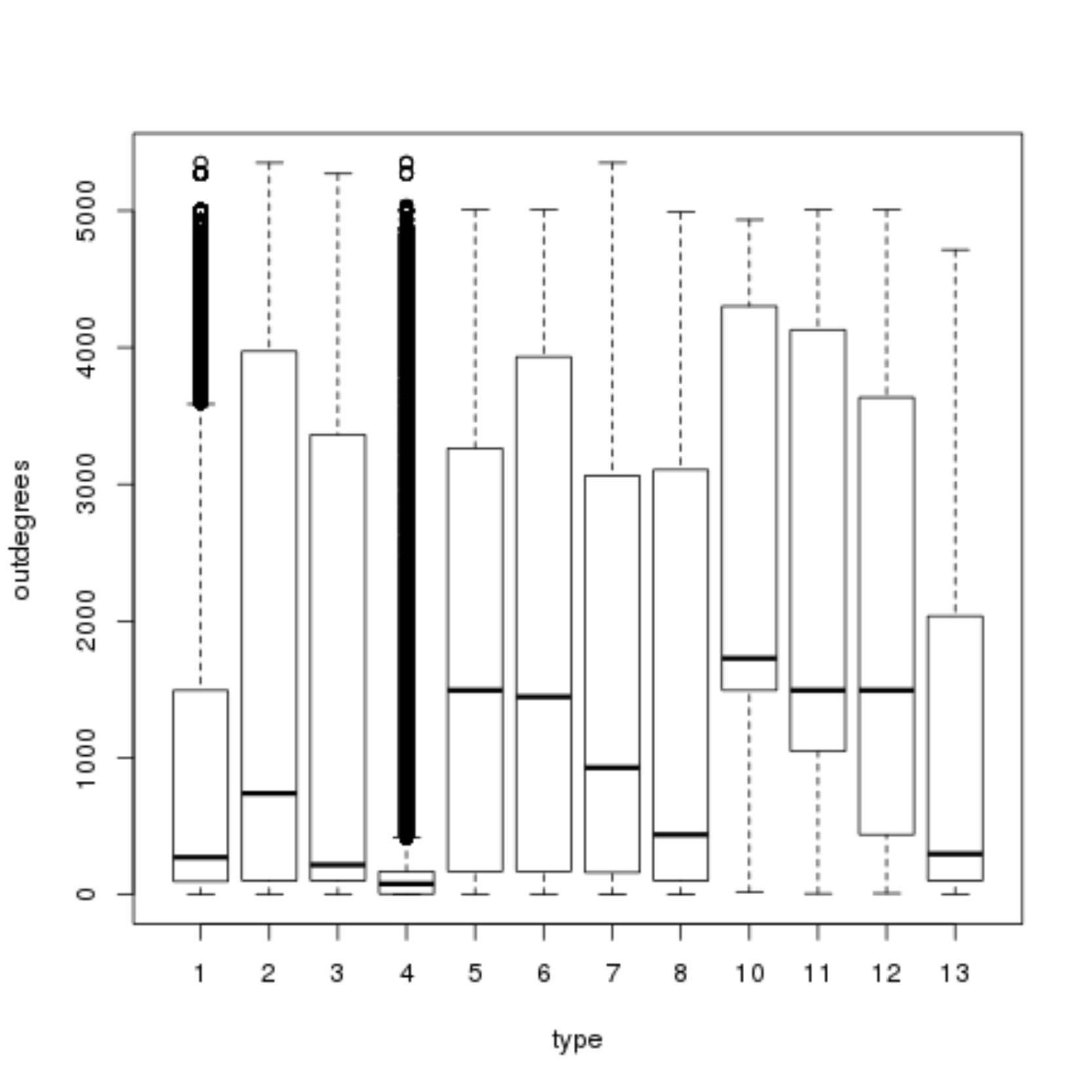}
\includegraphics[width=.32\linewidth,angle=0]{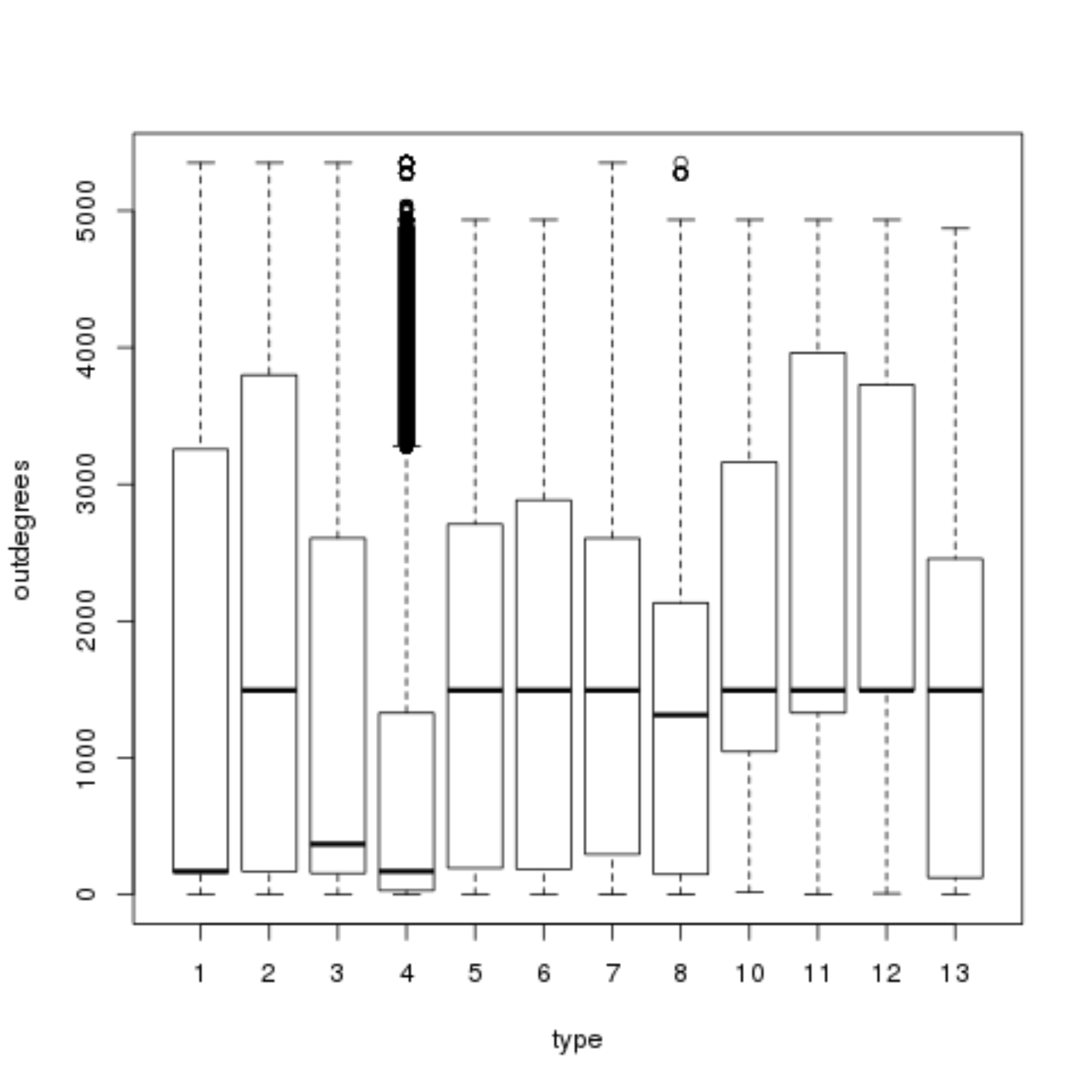}
\caption{\emph{Left:} Out-degrees of the triangle types that win edges over time.
\emph{Middle:} Out-degrees of the triangle types that loose edges over time.
\emph{Right:} Out-degrees of the triangle types that are stable over time.}
\label{fig:outdegree-win}
\end{figure*}

\fref{fig:outdegree-win} shows the out-degrees of nodes participating in a
certain triangle type: \fref{fig:outdegree-win} \emph{(top left)}
studies triangles that \emph{win} edges over time, i.e., transition into a
stronger connected type, \fref{fig:outdegree-win} \emph{(top right)} studies triangles that loose
edges over time, and \fref{fig:outdegree-win} \emph{(bottom)} shows the node degrees per triangle
type for all triangles that stay \emph{the same type} over time.

%
%
%
%
%
%

\begin{table*}[t]
\centering
\caption{Timezone-Neighbors}\label{tab:tria-tz-neigh}
\tabcolsep2mm
\begin{tabular}{lrrrrrrrrrrrrr}
\toprule
  Data   &  1  &  2 & 3 & 4 & 5 & 6 & 7 & 8 & 9 & 10 & 11 & 12 & 13 \\
\midrule
\# of triangles within three neighboring timezones & 417607 & 63132 & 234476 & 774584 & 9883 & 30913 & 40682 & 20795 & 2 & 80 & 1215 & 826 & 515 \\
percent of triangles within three timezones & 30.91 & 13.55 & 47.77 & 13.04 & 5.20 & 52.13 & 14.75 & 26.84 & 9.53 & 1.66 & 2.00 & 4.62 & 26.17 \\
percent of triangles within one timezone & 22.92 & 8.53 & 39.40 & 6.43 & 3.88 & 48.73 & 10.18 & 22.17 & 9.53 & 1.06 & 1.25 & 3.58 & 23.12 \\
types in general (percent of whole set) & 15.12 & 5.22 & 5.49 & 66.48 & 2.13 & 0.66 & 3.09 & 0.87 & 0.00024 & 0.05 & 0.68 & 0.20 & 0.02 \\
\bottomrule
\end{tabular}
\end{table*}

\begin{figure}[h]
\centering
\includegraphics[width=1.0\columnwidth,angle=0]{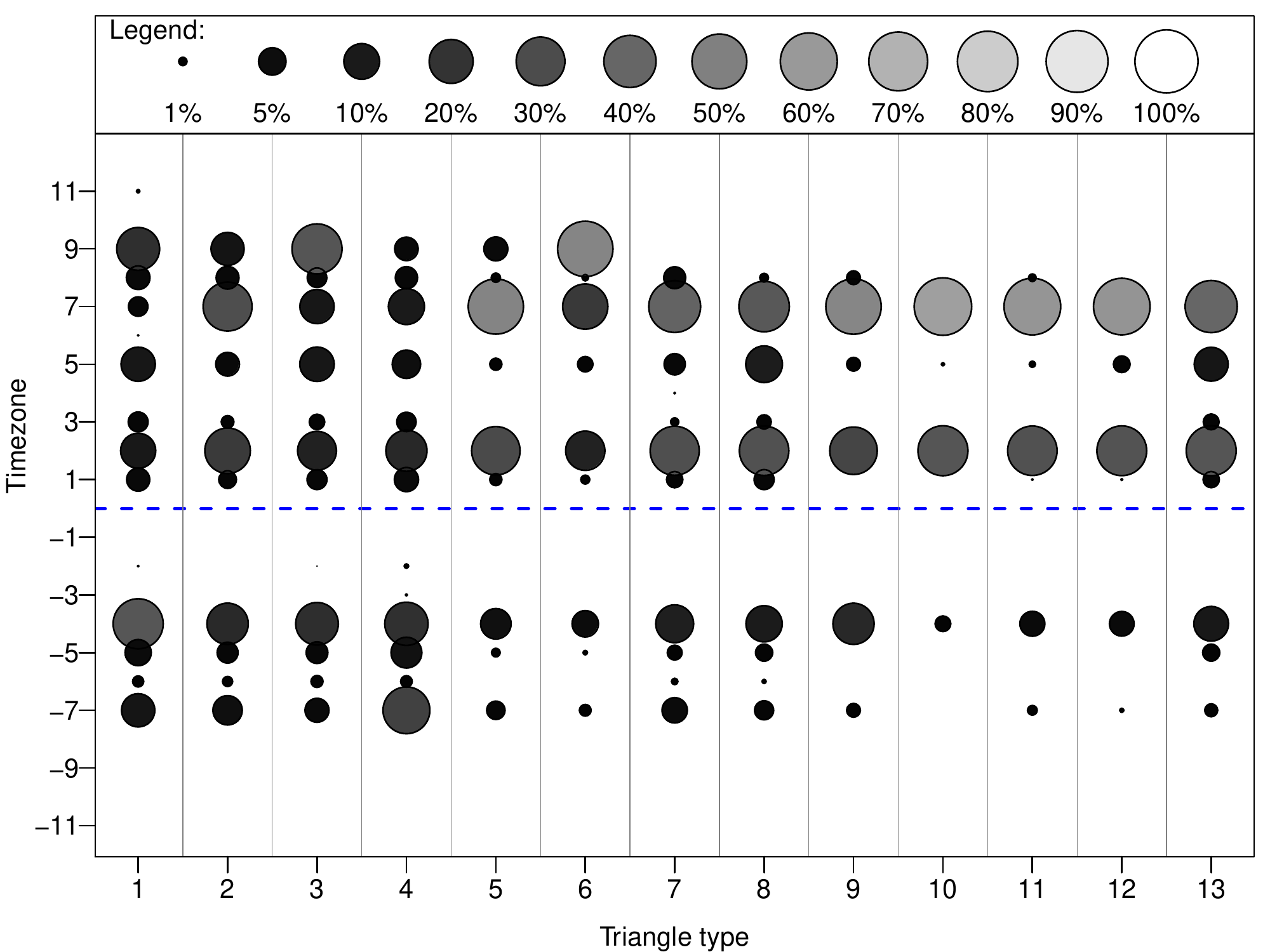}
\caption{Timezones in Triangles: timezones that appear in triangles by type}
\label{fig:tritimeheat}
\end{figure}

\subsection{Geographic Distribution}

To complement and extend our topological study, we also crawled the publicly available user profile data,
focusing mainly on geographic locations (taken from $S_4$). Based on this information, we can provide evidence
for our hypothesis that symmetric links are more likely to describe person relationships while asymmetric links
describe a follower or ``news-reader'' relationship (see Section~\ref{sec:data}).

\begin{figure}
\centering
\includegraphics[width=1.0\columnwidth,angle=0]{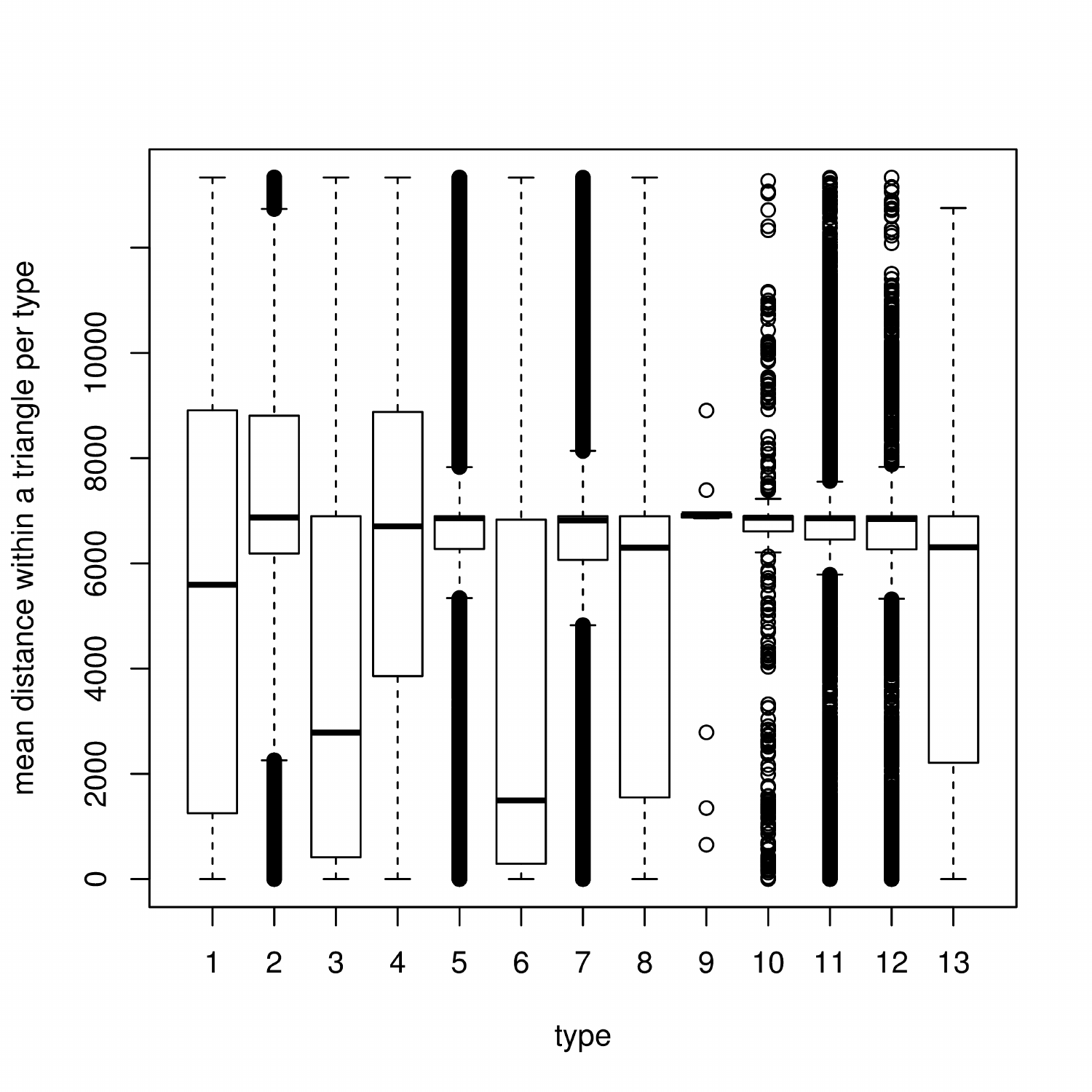}
\caption{Distances in triangles per type}
\label{fig:tridistptype}
\end{figure}

\begin{figure}
\centering
\includegraphics[width=1.0\columnwidth,angle=0]{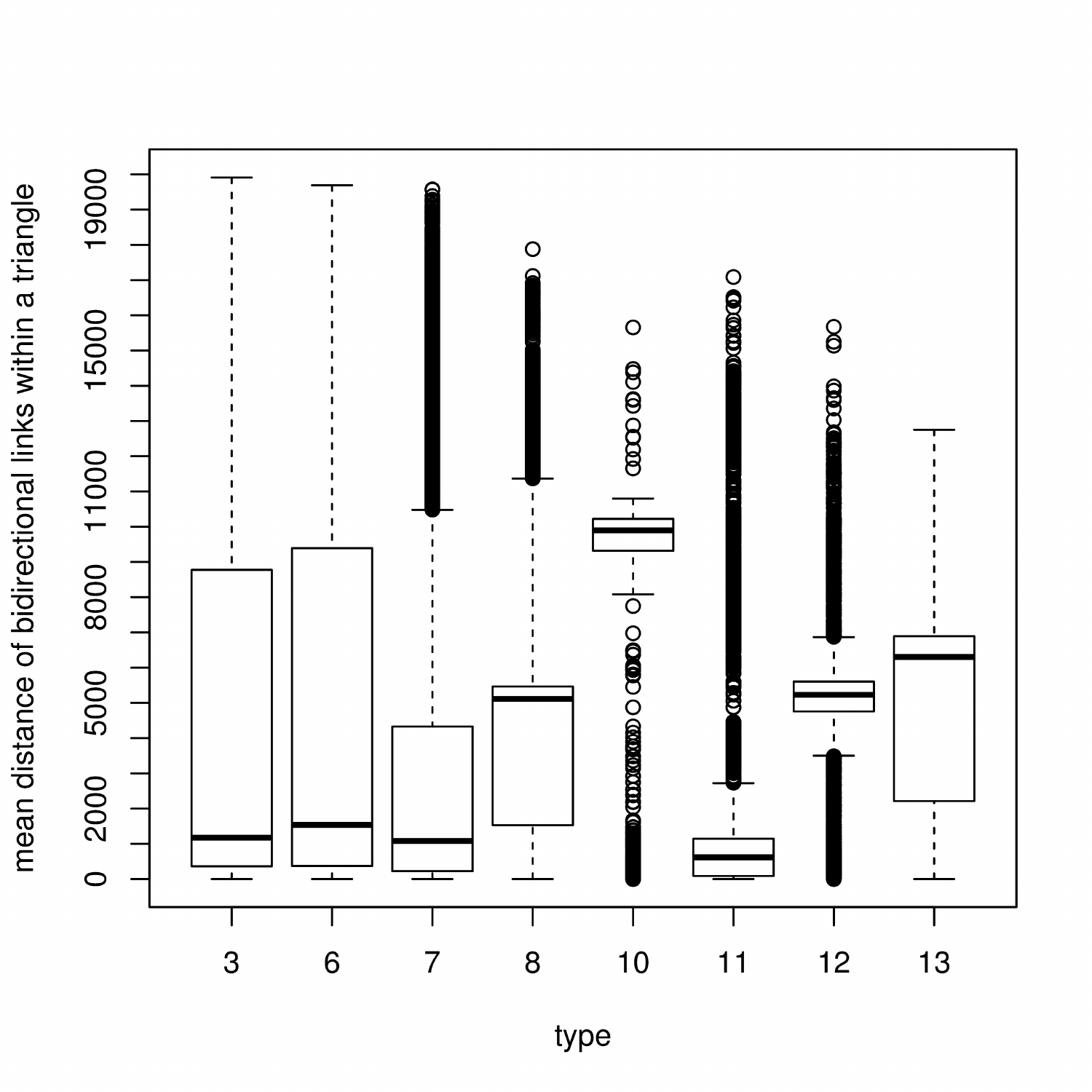}
\caption{Distances in triangles: mean distance of bidirectional links (triangle types without a bidirectional link are excluded).}
\label{fig:tridistbidir}
\end{figure}

\fref{fig:tridistptype} shows the mean of the distances per triangle,
grouped by motif types, and \fref{fig:tridistbidir} plots the mean distance of bidirectional links in triangles. We observe that
the mean distances of bidirectional links are much shorter, indicating that bidirectional links really seem
to be among people who have a personal contact, i.e., know each other. When focusing on the average distances
in triangles, the impact of one-directional links is quite large, making the overall distance more uniform
despite the naturally large variance.

\begin{figure}[h]
\centering
\includegraphics[width=1.0\columnwidth,angle=0]{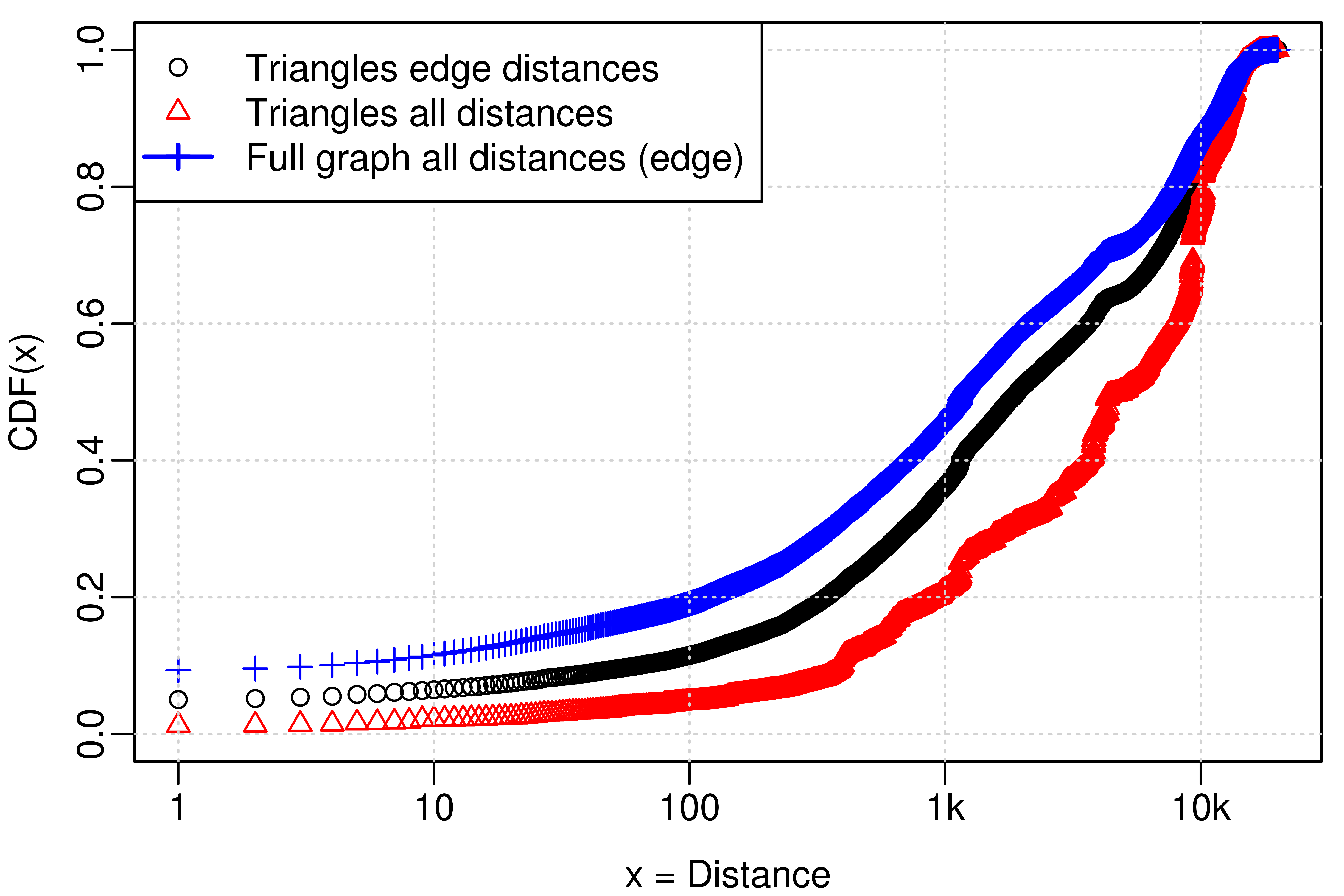}
\caption{Distances in triangles vs.~original graph}
\label{fig:dist-compare}
\end{figure}

For comparison purposes, \fref{fig:dist-compare} shows the CDF of the distances between the users for all distances we know, for the edges in the triangles, and for the users within a triangle. We calculate the distance between the users in a triangle even if there is no edge between those users. This is motivated by our assumption that all users in a triangle are somehow related to each other (at least transitively), so the distance of all participants matters. This figure highlights the bias on the distribution by accounting for these additional distances.

Finally, we also consider the different timezones users are located in. \fref{fig:tritimeheat}
gives an overview of the time zones of \gplus users in general, and \tref{tab:tria-tz-neigh}
studies on the number of triangles with users which are less than three time zones away from each other.
Looking at the distances in km, we can see now that the users in triangles of Type~3 and~6 are closest:
around half of the triangles of that type are within 2 timezones. For Type~8 and~13, this is only a third
of all triangles.

\section{Discussion}\label{sec:discussion}

We now discuss the shortcomings of relying on snapshots, especially how they are related to the challenges
that have to be overcome to observe the motif dynamics within an online social network such as Google$+$.

Relying on snapshots has the following limitations:\begin{enumerate}
\item[(L1)] Data provides limited resolution over time.
\item[(L2)] A snapshot does not describe a single point in time, but crawling a full snapshot does not scale and took almost a day.
\item[(L3)] All triangles could not be studied due to the combinatorial complexity of the triangles, only a random sample.
\end{enumerate}

One implication of Limitation~(L1) is that the motif dynamics will be underestimated: a user triple can be of the same triangle type in two snapshots, but may have gone through a sequence of changes
between---unnoticed. We believe that the fraction of changes we overlooked this way is relatively small, but in general, our results on the change rate must be understood as a conservative \emph{lower bound}.

Also Limitation~(L2) comes with certain implications. Generally, snapshots spanning longer time periods
cannot be used to study the causality of certain interactions. However, we in this paper did not
study such direct or causal interactions, hence we argue that the assumption is less critical.

\begin{figure}
\centering
\includegraphics[width=.49\linewidth,angle=0]{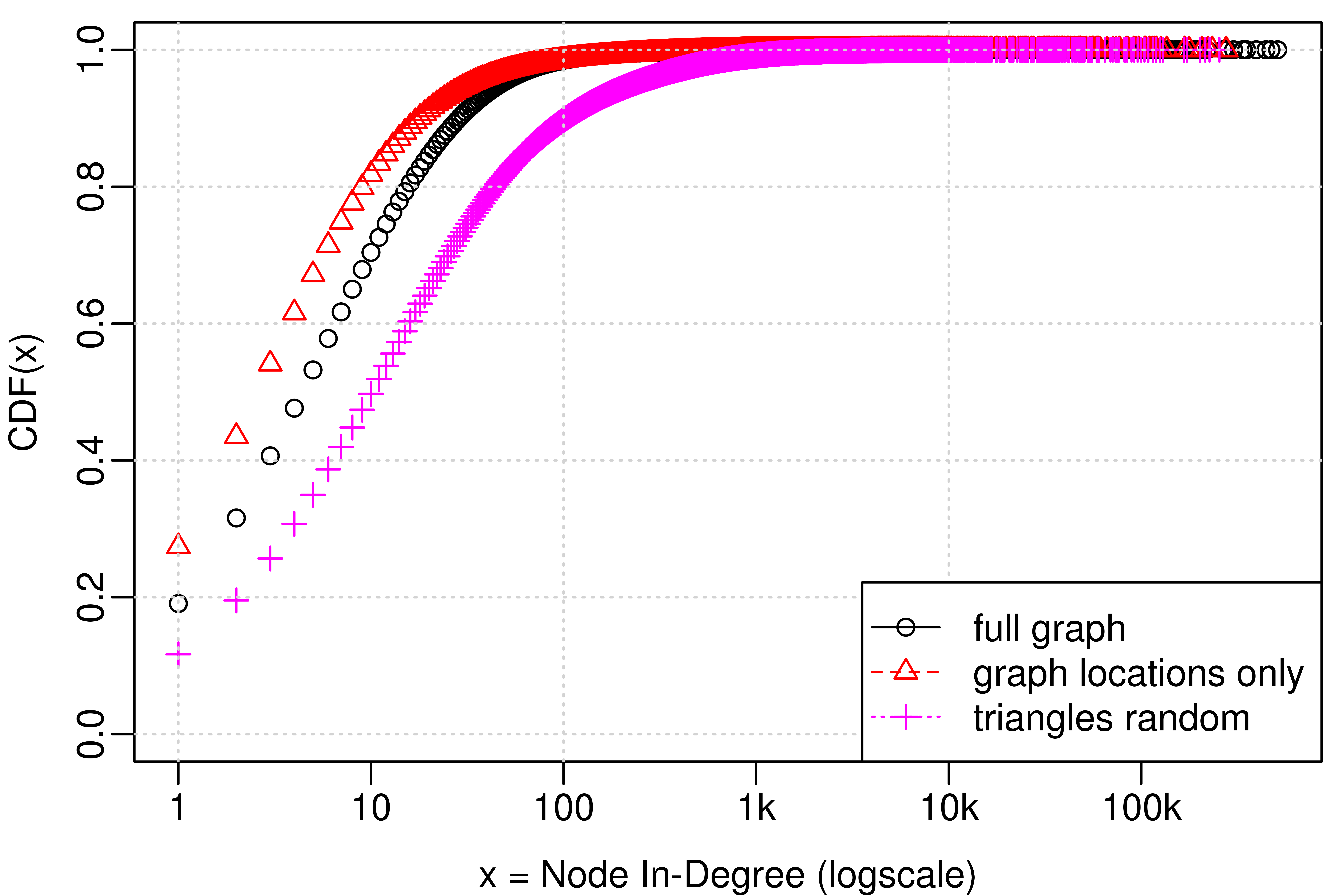}
\includegraphics[width=.49\linewidth,angle=0]{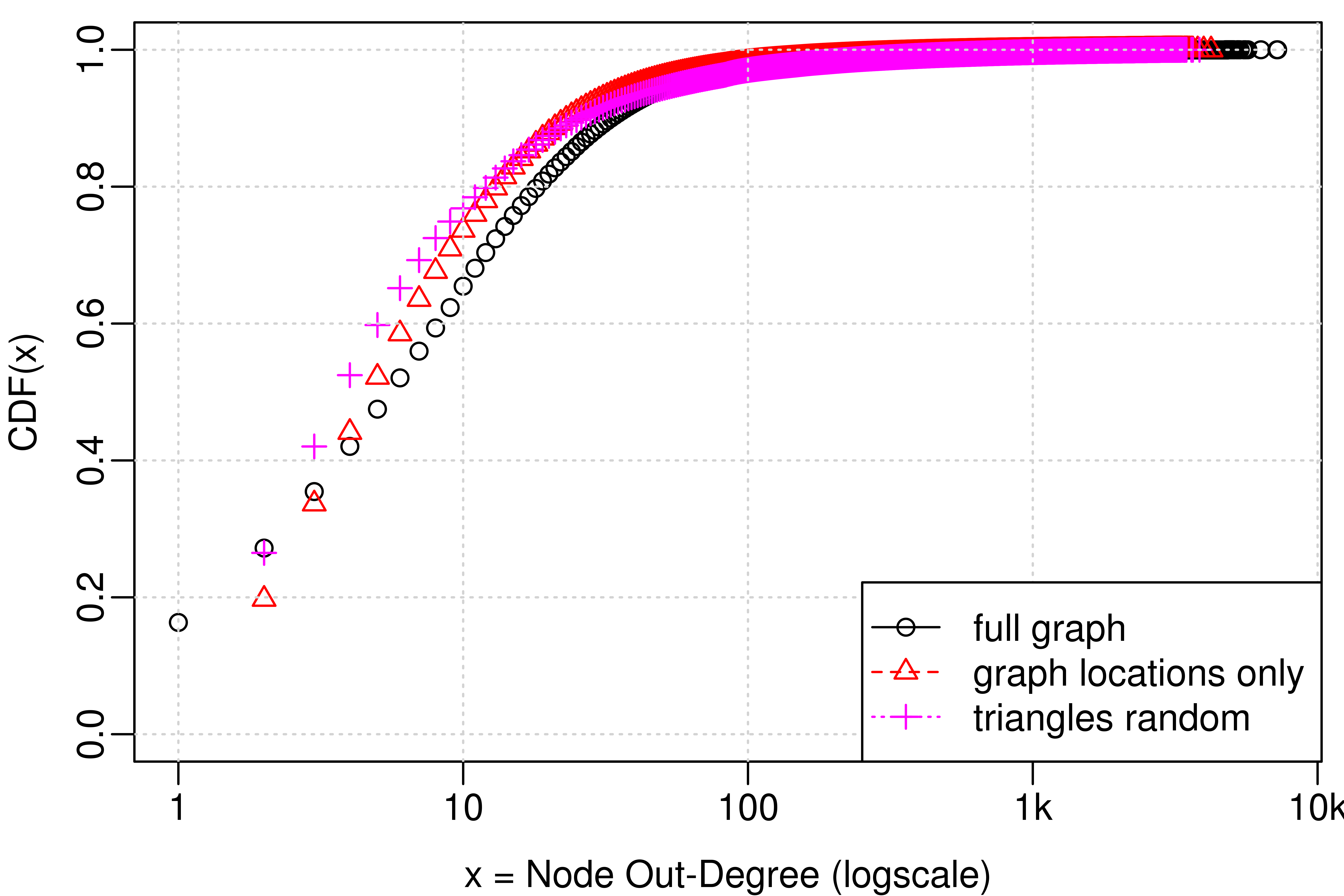}
\includegraphics[width=.49\linewidth,angle=0]{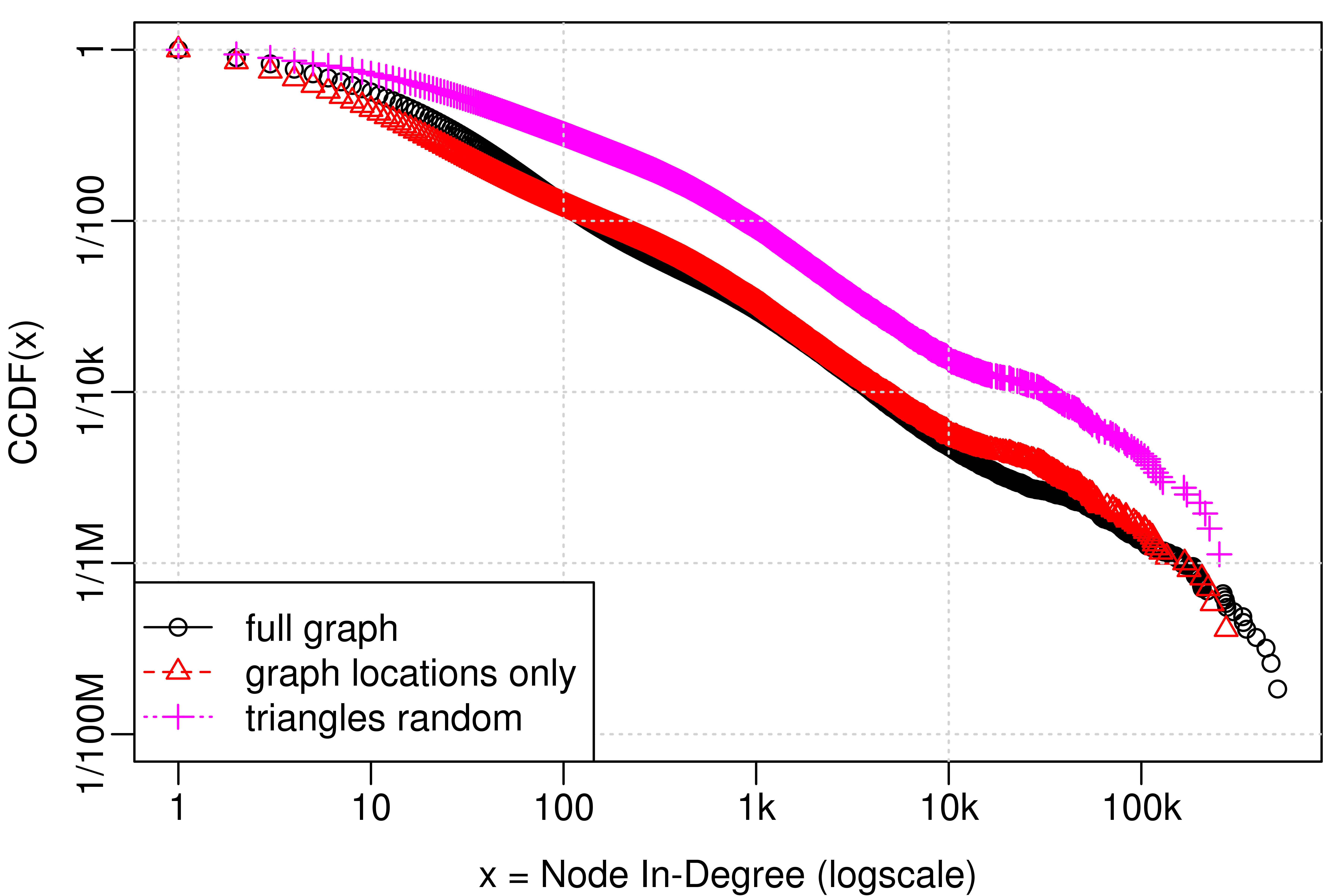}
\includegraphics[width=.49\linewidth,angle=0]{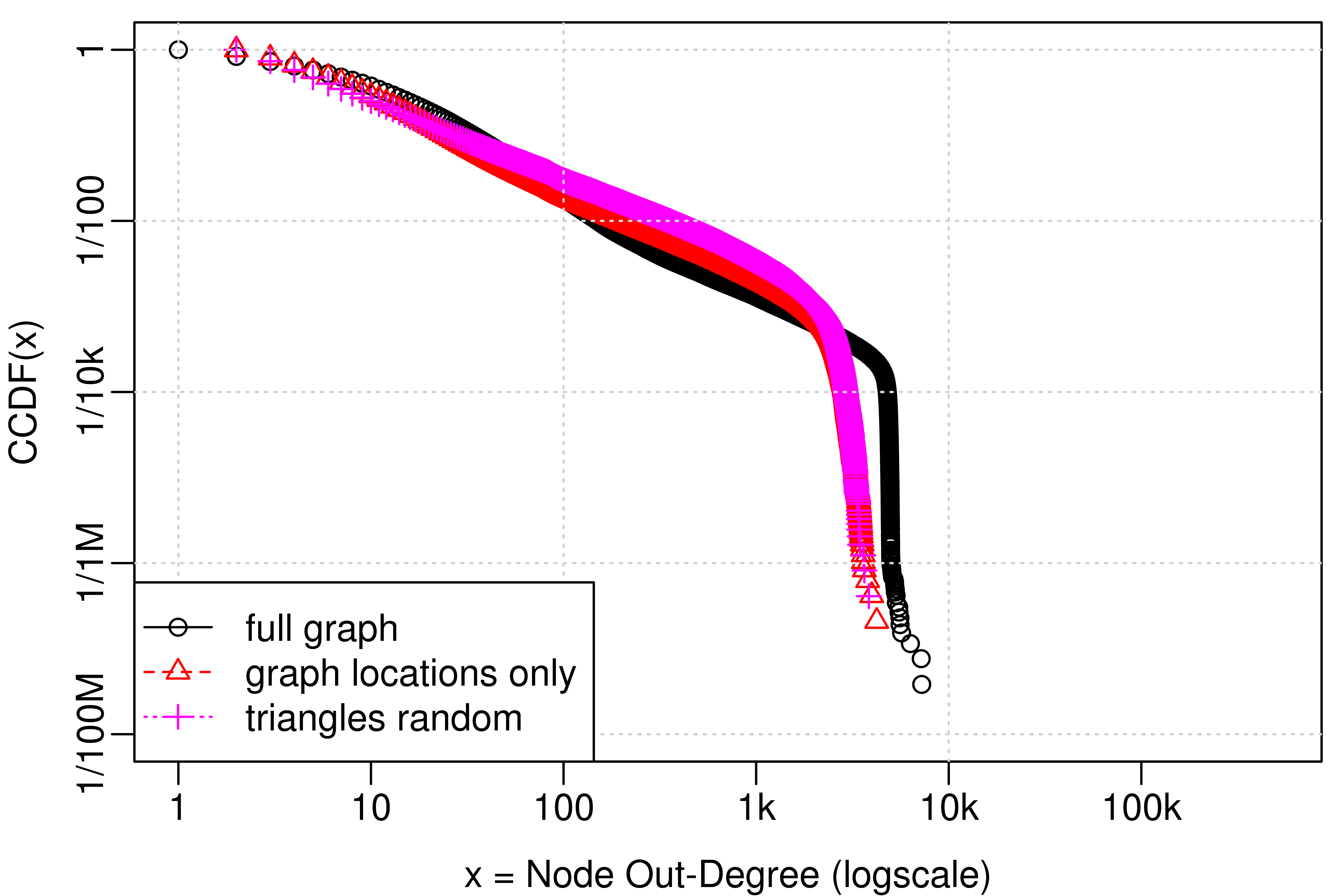}
\caption{\emph{Top left:} CDF of the in-degree distribution for the full Google$+$ graph compared to the graph spanned by the triangles;
\emph{Top right:} CDF of the out-degree distribution for the full Google$+$ graph compared to the graph spanned by the triangles.
\emph{Bottom left:} CCDF of the in-degree distribution for the full Google$+$ graph compared to the graph spanned by the triangles.
\emph{Bottom right:} CCDF of the out-degree distribution for the full Google$+$ graph compared to the graph spanned by the triangles.}
\label{fig:indegree-dist}
\end{figure}

Limitation~(L3) however is important: the focus on locations as well as the sampling process comes with a certain bias.
In the following, we will provide evidence that while the sampling process does play a role in our plots, the general nature of the system is preserved.
\fref{fig:indegree-dist} shows the CDF and CCDFs for both in- and out-degree-distributions after each reduction and sampling step.
The differences in the CDF are obvious: the heads of the distributions differ in their nature,
although the absolute numbers are relatively small.
However, we argue that more relevant for our study are the \emph{tail distributions}, i.e.,
the CCDFs. Here, the different samples are more similar. Moreover, the differences
do not concern the general \emph{shape} of the curve, but are rather \emph{shifted}.

While this is good news and indicates that our sampling methodology does not influence the
qualitative results much, a more rigorous analysis, also of additional properties, needs to be
conducted.

Let us conclude with a remark on the interpretation of the triangles. Any large empirical study
as ours is bound to aggregate and ignore many important details. For example, in our case,
all we know about the users is some geographic location and their connectivity. Accordingly, restricting ourselves
to the 13 triangle types and giving users of the same type a common interpretation is problematic. Although maybe large geographic
distances, asymmetric links and high in-degrees may suggest that users do not know each other in person, this may not
be true in general.
Therefore, in this paper, we try to avoid the semantic interpretation of the different triangle types, and leave
this for future sociological studies. All we can offer is some statistical interpretations.

\section{Related Work}\label{sec:relwork}

Researchers have been fascinated by the topological structure and
the mechanisms leading to them for many years. While early works
focused on simple and static networks~\cite{erdos}, later models, e.g.,
based on preferential attachment~\cite{barabasi}, also shed light on
how new nodes join the network, resulting in characteristic graphs.
Nevertheless, today, only very little is known about the dynamics of
social networks. This is also partly due to the lack of good data,
which renders it difficult to come up with good methodologies for
evaluating, e.g., link prediction algorithms~\cite{linkprediction,yang:friendship}.

\textbf{Motifs and Triangles.} Graph structures are often characterized by the frequency of small patterns
called \emph{motifs}~\cite{motif,motifs,motifs-bioinf,Schreiber05frequencyconcepts}, and also known as \emph{graphlets}~\cite{graphlet},
and \emph{structural signatures}~\cite{signatures}. The efficient computation of more complex motifs
is of independent interested, and the reader is referred to the corresponding literature, e.g.,~\cite{motifs-bioinf}.

Our paper focuses on the most simple triangle motif, whose importance
has been observed in many contexts. For example, triangles are
of interest for the study of community detection algorithms~\cite{community},
and also the frequently studied \emph{clustering coefficient}~\cite{wattsstrogatz} is defined based on
triangles. The clustering coefficient has many applications~\cite{cite11,cite13,cite27};
to give just one example, the clustering coefficient has recently been interpreted as
a curvature~\cite{curvature} and it has been shown that connected regions
of high curvature on the WWW characterize similar topics.

Many existing concepts, such as the clustering coefficient, are based on undirected triangles. However, also directed triangles have already been proposed to compare
graphs, e.g., in~\cite{directed1,signatures,directed2}.
In~\cite{cikm-tri-deg}, Durak et al.~take a closer look at the structure
of the different triangles, and study
degree relations in networks. They find that
networks
coming from social and collaborative situations are dominated by homogeneous triangles, i.e., degrees of vertices in
a triangle are quite similar to each other. On the other
hand, information networks (e.g., web graphs) are dominated by heterogeneous triangles.

\textbf{OSNs and Google$+$.}
For a (historic) overview of OSNs, the reader is referred
to~\cite{osn-history}.
In~\cite{mislove-2007-socialnetworks}, the authors report on a large-scale measurement
study of the topological structure of Flickr, YouTube,
LiveJournal, and Orkut, and the paper confirms the power-law,
small-world, and scale-free properties of OSNs. A demographic
perspective is assumed in~\cite{mislove-2011-twitter}, where Mislove
et al.~investigate how representative Twitter users are of the
overall population. Ahn et al.~\cite{Ahn:2007:ATC:1242572.1242685} take a
look at the growth patterns and topological (degree-based) evolution
of OSNs (Cyworld, MySpace, and Orkut) and compare their results with
the ones in real-life social networks. They focus on the scaling
exponent of the degree distribution, and find that certain OSNs
encourage on-line activities that cannot be easily copied in real
life, through the degree correlation pattern. In~\cite{cha-twitter},
Cha et al.~compare three topological measures of influence
(in-degree, re-tweets, and mentions) based on a large crawl of the
Twitter OSN. Scellato
et al.~\cite{geo-social} analyse the annotated geo-location graphs
of BrightKite, FourSquare, LiveJournal and Twitter, based on
snowball sampling crawls. They find that friendship edges are fairly
distant geographically, and define a new metric, called node
locality, which captures how close all neighbors of a node are.
Gjoka et al.~\cite{randomwalk} study parallel relations between OSN
users, by conducting  multigraph measurements of Last.fm.

Not much is known about the Google$+$ network.
In~\cite{websci12google}, Schi\"oberg et al.~investigate
topological properties of the network (e.g., in- and out-degree distributions),
and also use geographic information obtained from user profiles, e.g.,
on the distribution of link lengths. Kairam et al.~\cite{Kairam:2012:TCS:2207676.2208552}
study the circle structure of Google$+$, and its ramifications for selective
sharing. Gonzalez et al.~\cite{Gonzalez:2013:GGD:2488388.2488431} study the
user activity during the first year, and show that the network is not
strongly clustered. Gong et al.~\cite{Gong:2012:ESN:2398776.2398792}
develop a generative model to reproduce the social structure of Google$+$.

%





\section{Conclusion}\label{sec:conclusion}

We understand our work as a first step to shed light onto the dynamic evolution of triangle relationships
in Online Social Networks, and in particular Google$+$. In contrast to much existing literature focusing
on a single snapshot or on the question how a new user joins the network initially (i.e., bootstraps),
our study also considers changes in the longer run; indeed, we observe that quite a large number of links
 also disappears over time, even during this year of fast growth.  Accordingly, we believe that our methodology
 and results give an interesting new perspective on the field, and also have implications, e.g., on link prediction.

However, while our study shows a high degree of motif churn, the results are still very conservative. In fact, we show that
any methodology based on a discrete set of snapshots is bound to underestimate the dynamics: there is
a large fraction of Type~0 triangles, i.e., triangles which only exist during a subset of the snapshots. A higher
resolution of the evolution over time is hence likely to increase the churn rate further.

{\footnotesize
\renewcommand{\baselinestretch}{.73}
\bibliographystyle{IEEEtran}
\bibliography{gplus}
}

\end{document}

%% file: utils.tex
\newcounter{tub}
\newcounter{nec}
\newcounter{qmul}
\newtheorem{theorem}{Theorem}[section]

\newtheorem{definition}[theorem]{Definition}

\newcommand{\xref}[1]{Section~\ref{#1}}
\newcommand{\cref}[1]{Chapter~\ref{#1}}

\newcommand{\fref}[1]{Figure~\ref{#1}}
\newcommand{\tref}[1]{Table~\ref{#1}}

\newcommand{\ie}{i.\,e., \@}
\newcommand{\eg}{e.\,g., \@}

\newcommand{\perc}{\,\%\xspace}

\usepackage[normalem]{ulem}

\newcounter{fn1}
\newcounter{fn2}
\newcounter{fn3}
\newcounter{fn4}
\newcounter{fn5}
\newcommand{\gplus}{Google+\xspace}

\newcommand{\btmr}[1]{\multirow{2}{*}[-1mm]{#1}}

\newcommand{\tria}{\includegraphics[height=8pt]{tri-1doris}}
\newcommand{\trib}{\includegraphics[height=8pt]{tri-2doris}}
\newcommand{\tric}{\includegraphics[height=8pt]{tri-3doris}}
\newcommand{\trid}{\includegraphics[height=8pt]{tri-4doris}}
\newcommand{\trie}{\includegraphics[height=8pt]{tri-5doris}}
\newcommand{\trif}{\includegraphics[height=8pt]{tri-6doris}}
\newcommand{\trig}{\includegraphics[height=8pt]{tri-7doris}}
\newcommand{\trih}{\includegraphics[height=8pt]{tri-8doris}}
\newcommand{\trii}{\includegraphics[height=8pt]{tri-9doris}}
\newcommand{\trij}{\includegraphics[height=8pt]{tri-10doris}}
\newcommand{\trik}{\includegraphics[height=8pt]{tri-11doris}}
\newcommand{\tril}{\includegraphics[height=8pt]{tri-12doris}}
\newcommand{\trim}{\includegraphics[height=8pt]{tri-13doris}}